\documentclass{PoS}

\title{Prospects for Supersymmetry at the LHC \& Beyond}

\ShortTitle{Prospects for Supersymmetry}

\author{\speaker{John Ellis}\thanks{I thank members of the MasterCode Collaboration: 
{\tt http://mastercode.web.cern.ch/mastercode/} and other collaborators on work described here, which
has been supported in part by the European Research Council via the Advanced Investigator
Grant 267352 and by the UK STFC via the research grant ST/L000326/1.}\\
        Theoretical Particle Physics and Cosmology Group, Department of Physics, \\
King's College London, Strand, London WC2R 2LS, U.K; \\
Theory Division, Physics Department, CERN, CH 1211 Geneva 23, Switzerland\\
        E-mail: \email{John.Ellis@cern.ch}\\
        ~\\
      {\tt ~~~~~~~~~~KCL-PH-TH/2015-48, LCTS/2015-36, CERN-PH-TH/2015-250}  }

\abstract{Run 1 of the LHC has provided three new motivations for supersymmetry: the need to stabilize
the electroweak vacuum, the mass of the Higgs boson, and the fact that its couplings are Standard Model-like (so far).
The prospects for discovering (and measuring) supersymmetry during future runs of the LHC are discussed in the
frameworks of the constrained MSSM (CMSSM), models with non-universal soft supersymmetry-breaking
contributions to Higgs masses (NUHM1,2) and the phenomenological MSSM with 10 arbitrary soft supersymmetry-breaking
parameters (pMSSM10). In addition to the classic searches for missing transverse energy, searches for long-lived charged
sparticles may also be promising. If supersymmetry does show up at the LHC, there are good prospects for
measurements of the spectrum that can be compared with the indirect indications from other experiments.
On the other hand, a higher-energy future circular proton-proton collider may be the best option for discovering
supersymmetry if it does not appear at the LHC.}

\FullConference{18th International Conference From the Planck Scale to the Electroweak Scale \\
		 25-29 May 2015\\
		 Ioannina, Greece}

\begin{document}

\section{Introduction}

For lovers of supersymmetry, the good news from Run 1 of the LHC has been that there has been no
evidence for any rival scenario for physics beyond the Standard Model (SM). The hapless fate
of Higgsless models has been sealed, and Run~1 of the LHC found no signs of extra dimensions or compositeness. On the other hand, the
bad news has been that there has also been no direct evidence for superymmetry. Nevertheless, I argue
that there have been three indirect pieces of evidence favouring supersymmetry: the measured masses of
the Higgs boson and the top quark suggest that the electroweak vacuum would be unstable within the SM
(whereas it could be stabilized by supersymmetry), the measured Higgs mass is within the range predicted by
simple supersymmetric models, and these also predicted successfully that its couplings would be similar to those in the SM.
Advocates of many alternatives should be more discouraged than we lovers of supersymmetry. We lovers of supersymmetry
should redouble our ardour, renewing searches for missing transverse energy (MET) and looking for possible long-lived
charged sparticles. This talks presents my personal take on the remaining prospects for discovering (and measuring) supersymmetry at Run~2
and future runs of of the LHC, as well as previewing the prospects for future circular colliders.

\section{The Collapse of the Electroweak Vacuum, and How to Avert it}

\begin{figure}[htb]
\centering
\includegraphics[height=2.5in]{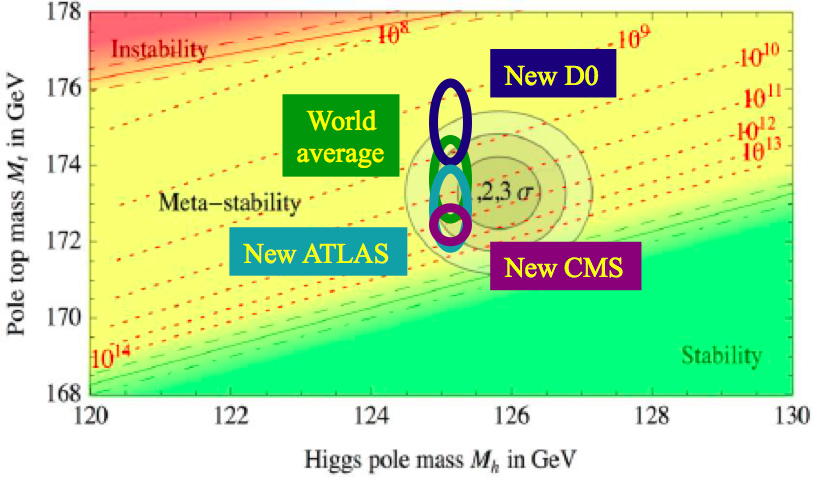}
\caption{\it The $(m_H, m_t)$ plane showing regions of vacuum (meta/in)stability~\protect\cite{Buttazzo} and the 68\% CL
regions favoured by LHC measurements of $m_H$~\protect\cite{mH} and the 2014 world average
measurement of $m_t$~\protect\cite{WAmt}, as well as more recent measurements~\protect\cite{CMSmt}.}
\label{fig:mHmt}
\end{figure}

In the SM, the quartic Higgs self-coupling $\lambda$ is renormalized by itself and by
its Yukawa coupling to the top quark. The latter renormalization is negative, tending to
drive $\lambda < 0$ and induce vacuum instability at a scale $\Lambda_I$, which was estimated in~\cite{Buttazzo} to be
\begin{equation}
\log_{10} \left( \frac{\Lambda_I}{\rm GeV} \right) \; = \; 11.3 + 1.0 \left(\frac{m_H}{\rm GeV} - 126 \right)
- 1.2 \left( \frac{m_t}{\rm GeV} - 173.10 \right) + 0.4 \left( \frac{\alpha_s(m_Z) - 0.1184}{0.0007} \right) \, .
\label{LambdaI}
\end{equation}
Uisng the central values of the current world averages of $m_t = 173.34 \pm 0.76$~GeV~\cite{WAmt}, $m_H = 125.09 \pm 0.24$~GeV~\cite{mH}
and $\alpha_s(M_Z) = 0.1184 \pm 0.0006$~\cite{PDG},
it seems that the current electroweak vacuum is metastable, as seen in Fig.~\ref{fig:mHmt}, and
adding the uncertainties in quadrature in (\ref{LambdaI}) one estimates
\begin{equation}
\log_{10} \left( \frac{\Lambda_I}{\rm GeV} \right) \; = \; 11.1 \pm 1.3 \, .
\label{LambdaIvalue}
\end{equation}
Taken in isolation, 
the new CMS measurement of $m_t =  172.44 \pm 0.49$~GeV~\cite{CMSmt} would modify this to $\log_{10} (\Lambda_I/{\rm GeV}) = 11.6 \pm 0.7$.
This would be changed to $\log_{10} (\Lambda_I/{\rm GeV}) = 11.2 \pm 0.9$ if one also took $\alpha_s(M_Z) = 0.1177 \pm 0.0013$
as might be suggested by a re-evaluation of lattice estimates~\cite{Siggi}. If there were no physics beyond the SM, Higgs field values above $\Lambda_I$ 
would have a lower energy than the current SM vacuum, which is therefore unstable (though very long-lived).

\begin{figure}[htb]
\centering
\includegraphics[height=1.7in]{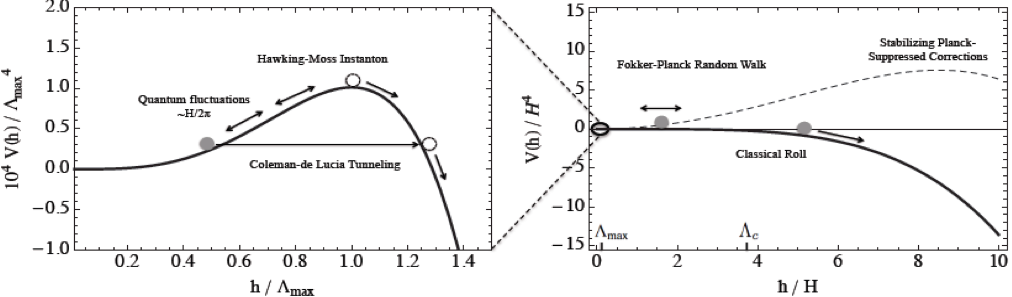}
\caption{\it In the early Universe there would have been (left panel) an enhanced probability for a transition
away from the neighbourhood of the electroweak vacuum towards large Higgs field values~\protect\cite{Hook}, and (right panel)
averting this catastrophe would have required new physics such as supersymmetry~\protect\cite{ER} or higher-dimensional
interactions with coefficients suppressed by powers of the Planck mass~\protect\cite{Sher}.}
\label{fig:Hook}
\end{figure}

But should one worry at all about this apparent instability? Certainly, we must hope for greater accuracy in
the experimental measurements of $m_t$ and $\alpha_s (M_Z)$ to establish definitively whether we are in
the region that would be meta/unstable within the SM. Also, since the lifetime for vacuum decay is probably
much longer than the age of the Universe, one may wonder whether its metastability is a relevant issue. I would answer `yes', for two reasons. 
1) If our vacuum is not the true one, why is our present vacuum energy apparently adjusted to a very small in natural units?
Moreover, calculations indicate that, if there is a lower-energy state out there, one would have expected most of
the Universe to have fallen into it thanks to the large Higgs field fluctuations in the
early Universe~\cite{Hook}, as illustrated in Fig.~\ref{fig:Hook}.
Averting this potential catastrophe requires
some new physics beyond the SM, which could be supersymmetry~\cite{ER},
though many other types of new physics, such as dimension-6 operators~\cite{Sher}, could also do the job.

\section{{\it Floreat} Supersymmetry, but which Model?}

As already mentioned, stabilizing the electroweak vacuum is just one of three new motivations for supersymmetry
provided by data from LHC Run~1, the others being its successful predictions that $m_h \lesssim 130$~GeV~\cite{susymH} and
that its couplings would be similar to those in the SM~\cite{EHOW}. In contrast, whilst the Higgs mass and couplings could be
accommodated within a generic composite model, this is not such an automatic prediction. These new motivations for
supersymmetry can be added to all the traditional ones including the naturalness of the mass hierarchy, grand
unification, its necessity within string theory and the fact that supersymmetric models contain a very plausible
dark matter candidate~\cite{EHNOS}. 

Of course, it is disappointing that supersymmetric particles did not appear at the LHC during Run~1, and this
certainly makes it a less complete solution to the naturalness problem. On the other hand, one's tolerance
for a certain amount of fine-tuning is very much a personal matter, and supersymmetry at any scale is less
unnatural than the SM without supersymmetry. In parallel, the LHC also did
not produce any experimental evidence for any of the other proposed ways of addressing the naturalness issue
and, while some new ideas are now emerging~\cite{McCullough}, none of them is yet as compelling as supersymmetry.
Therefore, I am staying on the supersymmetric ship, even if it is leaking and listing a bit.

Most experimental searches for supersymmetry have been interpreted within the minimal supersymmetric
extension of the SM (the MSSM), and I assume this framework in the following. This model conserves $R$
parity, and hence contains a suitable candidate for dark matter, which I also assume in the following. This
leads to the characteristic missing-transverse-energy (MET) signature of supersymmetry that has been the
basis for most LHC searches. Initially, these were often interpreted within the constrained MSSM (CMSSM)~\cite{MC9}
in which the soft supersymmetry-breaking contributions to the scalar masses $m_0$, the gaugino masses $m_{1/2}$
and the trilinear scalar couplings $A_0$ were each assumed to be universal. However, this framework
incorporates additional assumptions that may well not be valid. Natural flavour conservation would only
impose universality on the masses of sfermions with the same quantum numbers, grand unification within SU(5)
would allow different masses for sfermions in $\mathbf{\overline 5}$ and $\mathbf{10}$ representations,
and there is no obvious reason why the soft supersymmetry-breaking contributions to the masses of the two Higgs doublets
should be the same as for the sleptons and squarks. Hence one may consider models with one or two non-universal
Higgs mass parameters (NUHM1,2)~\cite{MC9,MC10}, and one may consider models in which no assumptions are made about the
soft supersymmetry-breaking parameters, which are teated as purely phenomenological parameters (the pMSSM)~\cite{MC11}.

\section{Global Fits}

In this Section I discuss some results from global fits to the CMSSM - which has 4 free parameters $m_0, m_{1/2}, A_0$
and $\tan \beta$, the NUHM1 - which also has an extra common Higgs mass parameter, the NUHM2 - which has two extra Higgs
mass parameters $m_{1,2}$, and the pMSSM10 - a variant of the pMSSM with 10 free parameters, including 3 gaugino masses, different squark
masses for the first/second and third generations, a common slepton mass, a common trilinear coupling $A_0$, $\mu$, $M_A$ and $\tan \beta$.
For each model, we sample the multi-dimensional parameter space using the {\tt MultiNest}
algorithm with, e.g., $1.2 \times 10^9$ points to sample the pMSSM10 parameter space~\cite{MC12}.
In addition to the Higgs mass, other LHC Higgs measurements and the negative results of LHC searches for supersymmetry, our global fits include many 
electroweak precision and flavour observables, $g_\mu - 2$, the dark matter density and upper limits on spin-independent
dark matter scattering. These measurements are used to construct a global $\chi^2$ function that we interpret using a
frequentist approach~\cite{Fittino,Bayes}.
 
\begin{figure*}[htb!]
\begin{center}
\resizebox{7cm}{!}{\includegraphics{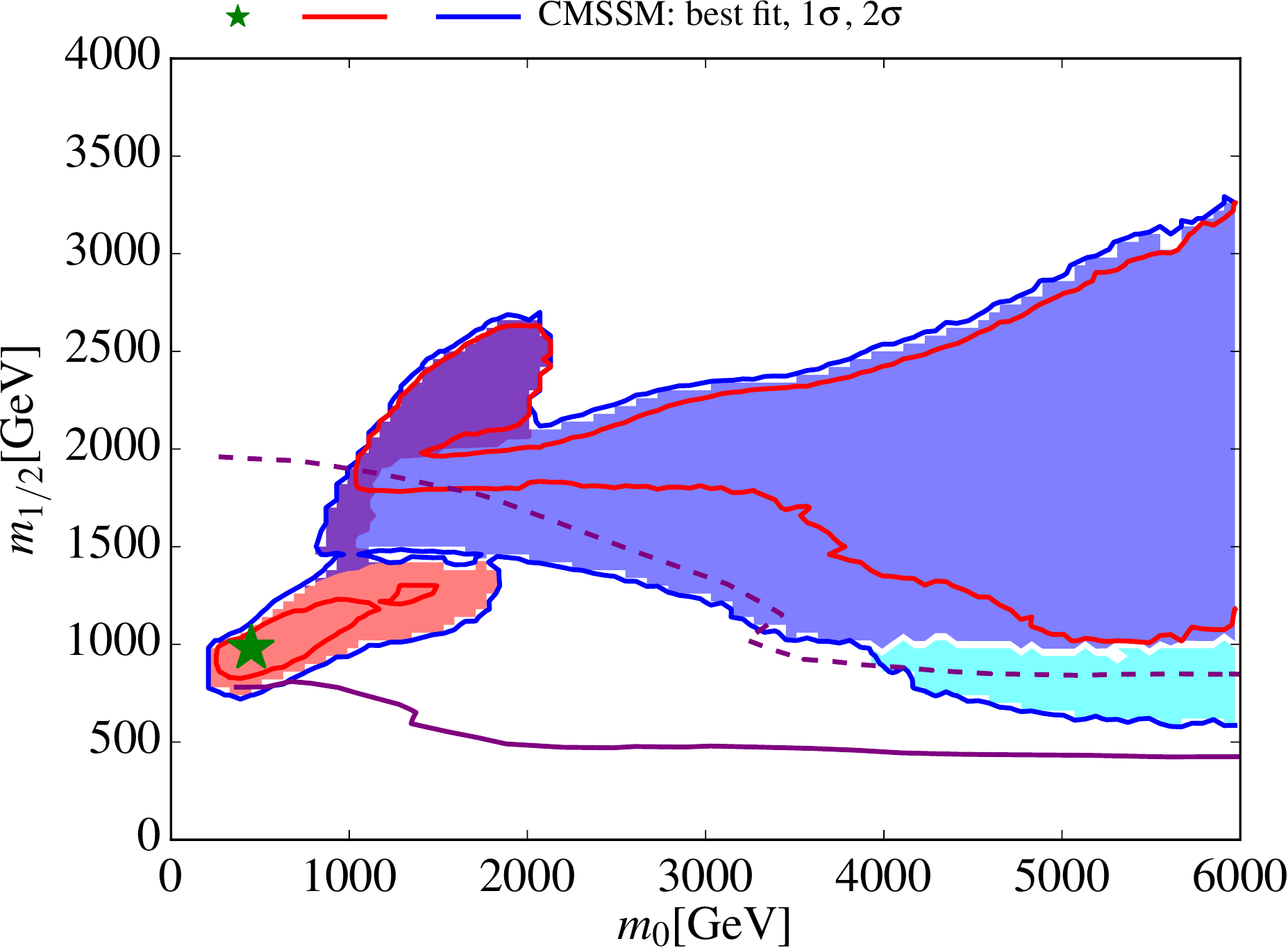}}
\resizebox{7cm}{!}{\includegraphics{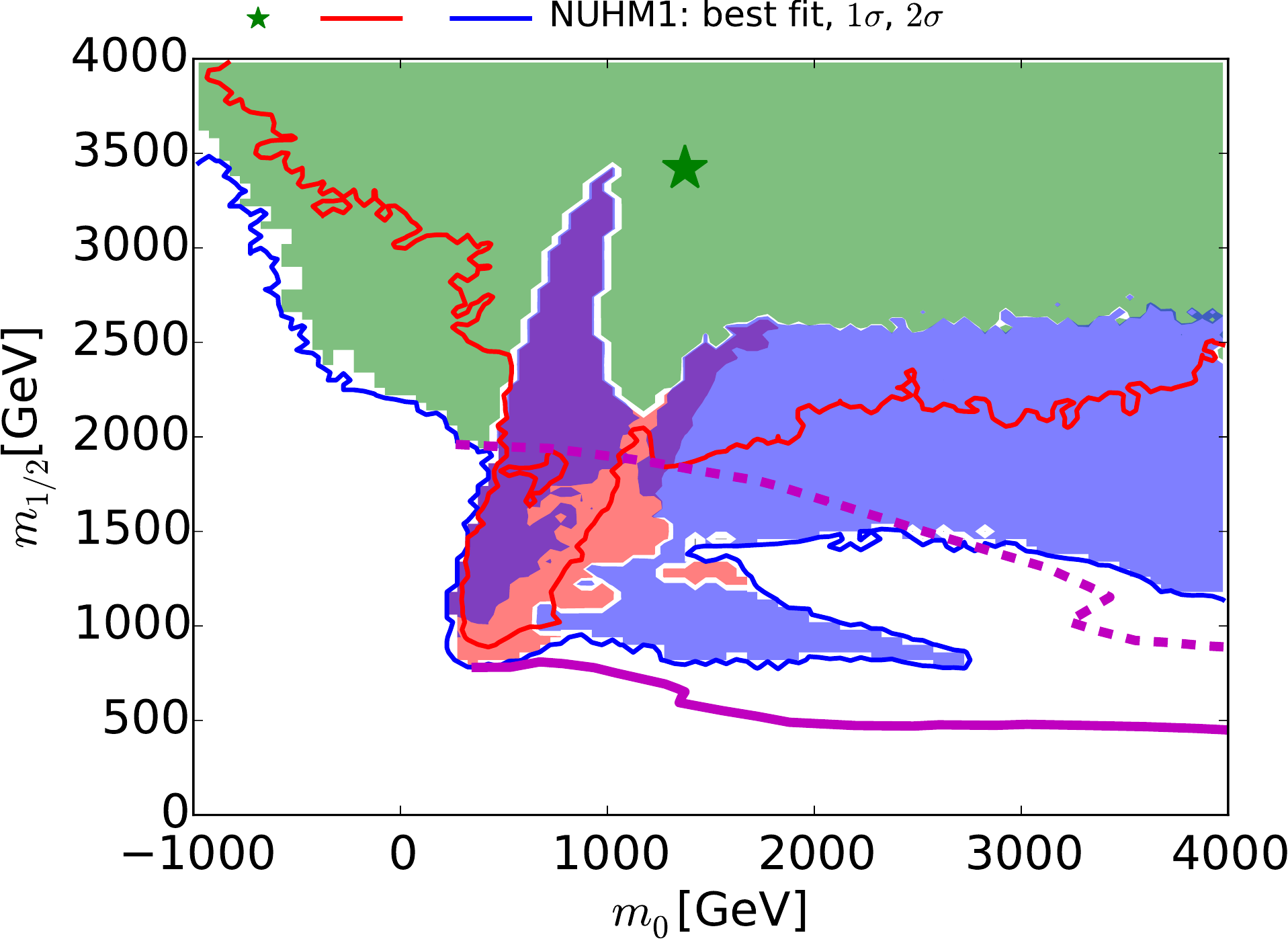}}\\
\resizebox{7cm}{!}{\includegraphics{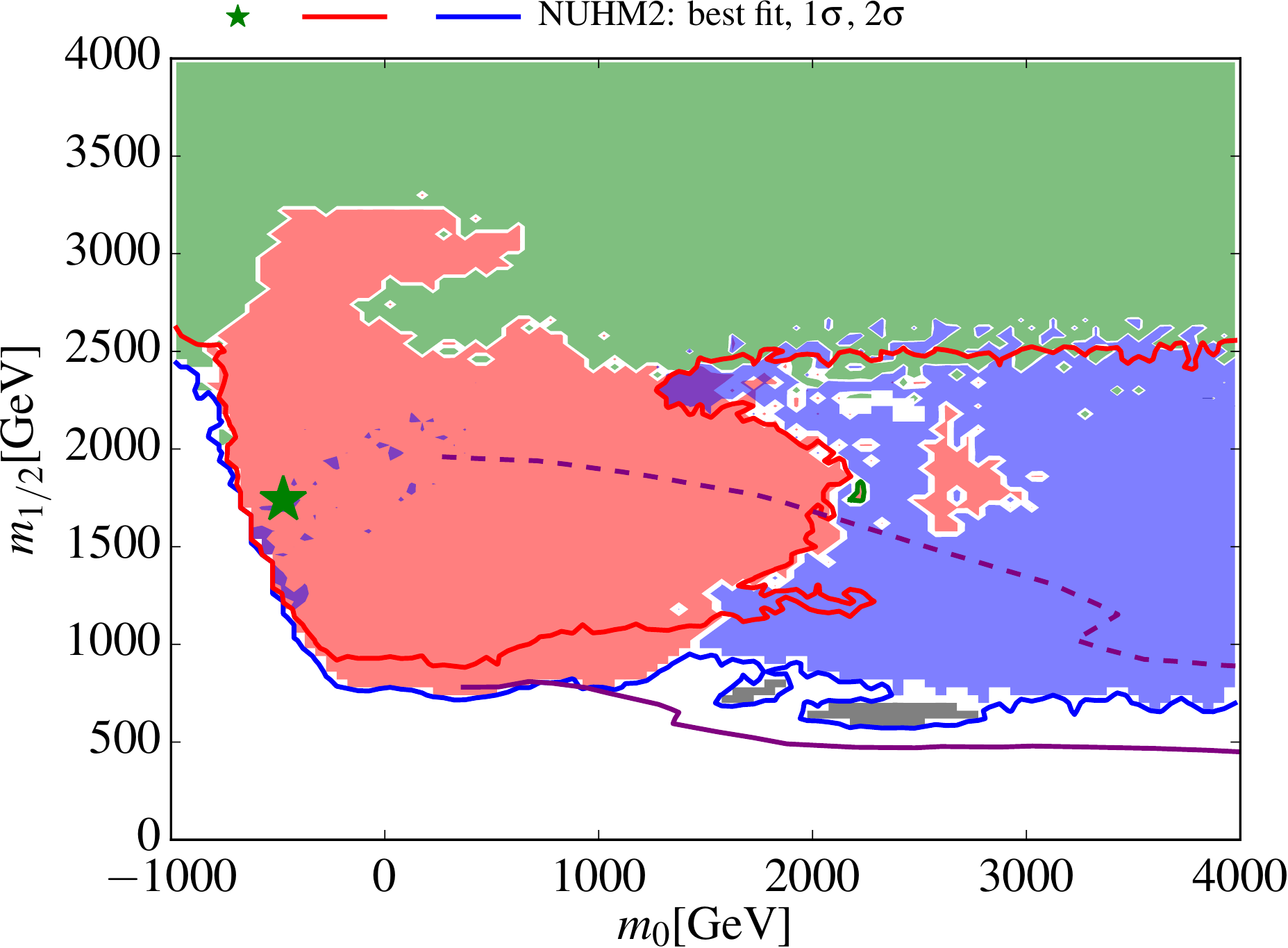}}
\resizebox{7cm}{!}{\includegraphics{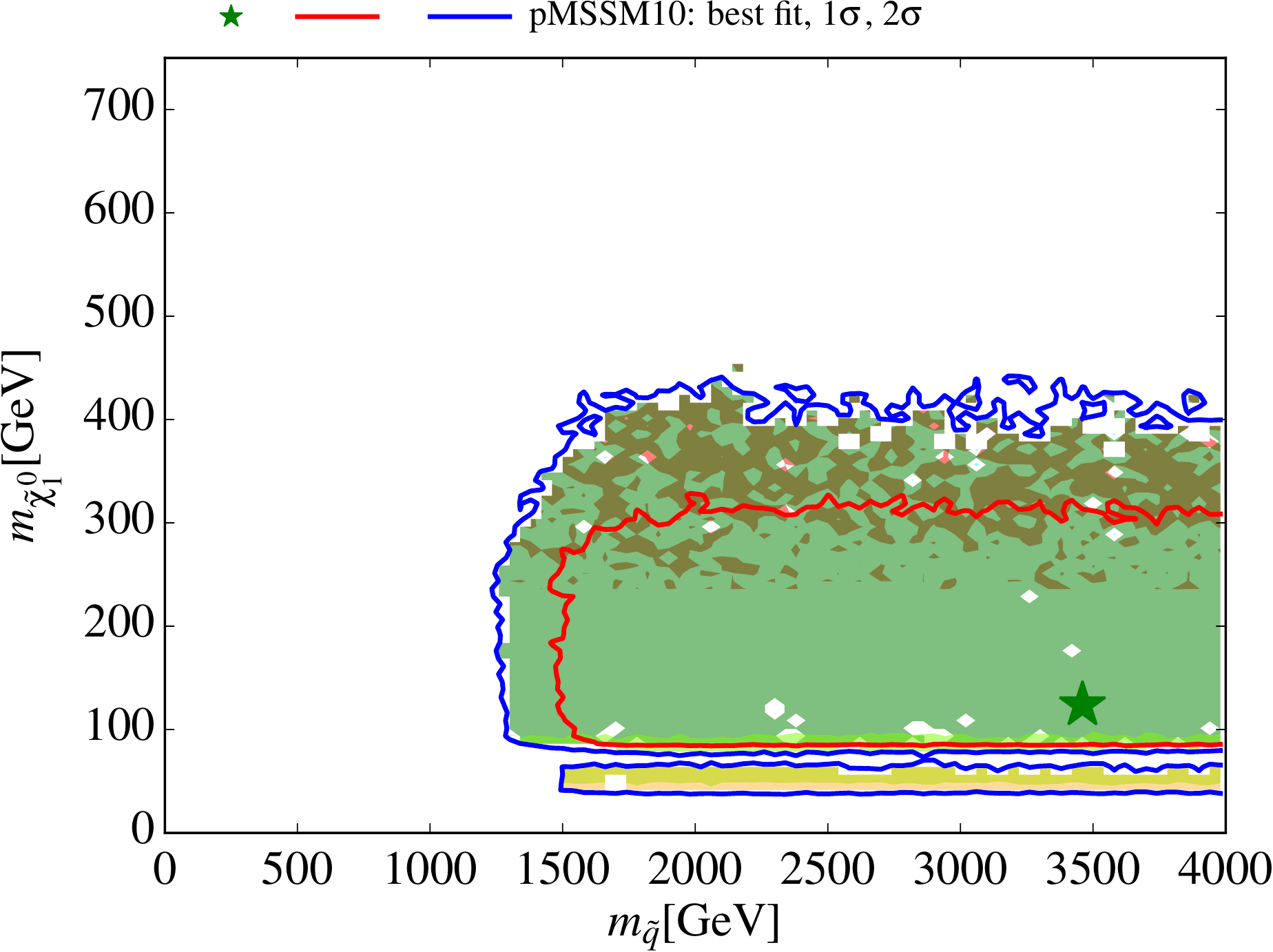}} \\
\vspace{0.2cm}
\resizebox{14cm}{!}{\includegraphics{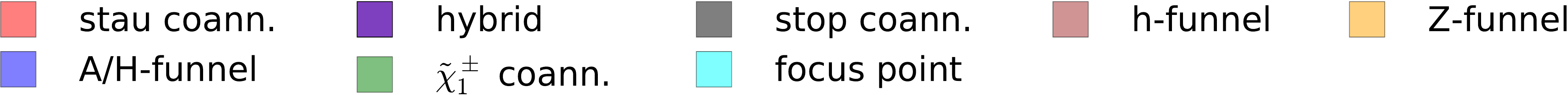}}
\end{center}
\vspace{-0.5cm}
\caption{\it The $(m_0, m_{1/2})$ planes in the CMSSM (upper left)~\protect\cite{MC9}, the NUHM1 (upper right)~\protect\cite{MC9} 
and the NUHM2 (lower left)~\protect\cite{MC10},
and the $(m_{\tilde q}, m_\chi)$ plane in the pMSSM10 (lower right panel)~\protect\cite{MC11}. Regions in which different mechanisms bring the
cold dark matter (DM) density into the allowed range are shaded as described in the legend~\protect\cite{MC12}. The red and blue
contours are the $\Delta \chi^2 = 2.30$ and $5.99$ contours found in global fits
to these models, the green stars indicate the best-fit points,
the solid purple contours
show the current LHC 95\% exclusions from MET searches, and the dashed purple contours show the prospective 5-$\sigma$
discovery reaches for MET searches at the LHC with 3000/fb at 14~TeV, which also correspond approximately to the 95\% CL exclusion
sensitivity with 300/fb at 14~TeV.}
\label{fig:m0m12}
\end{figure*}

Fig.~\ref{fig:m0m12} displays $(m_0, m_{1/2})$ planes for the CMSSM (upper left), the NUHM1 (upper right) and the NUHM2
(lower left), as well as the $(m_{\tilde g}, m_\chi)$ plane for the pMSSM10 (lower right panel). In each case, the best-fit point is indicated by a
green star, and the $\Delta \chi^2 = 2.30$ and $5.99$ contours (corresponding roughly to the 68 and 95\% CL boundaries)
are indicated by red and blue lines, respectively. The 95\% CL region is shaded according to the mechanism that is most important
for bringing the relic LSP density into the range of cold dark matter density favoured by Planck and other measurements,
with the colour coding illustrated above the figure caption~\cite{MC12}.

\begin{figure}[htb]
\centering
\includegraphics[height=4in]{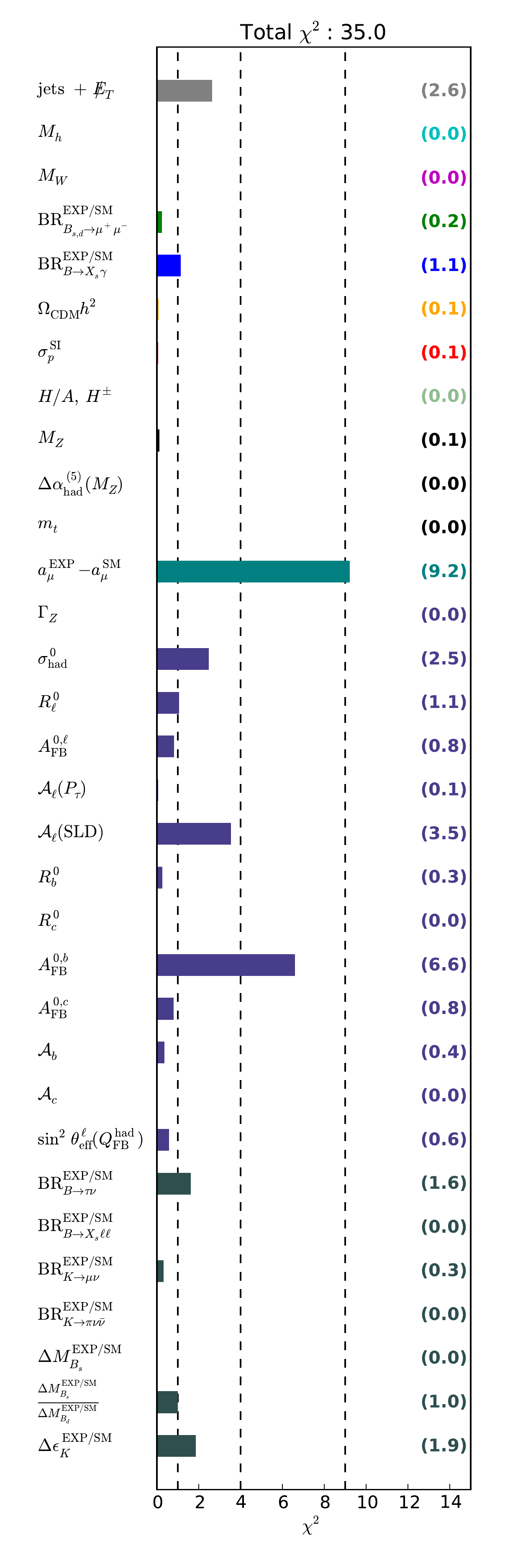}
\includegraphics[height=4in]{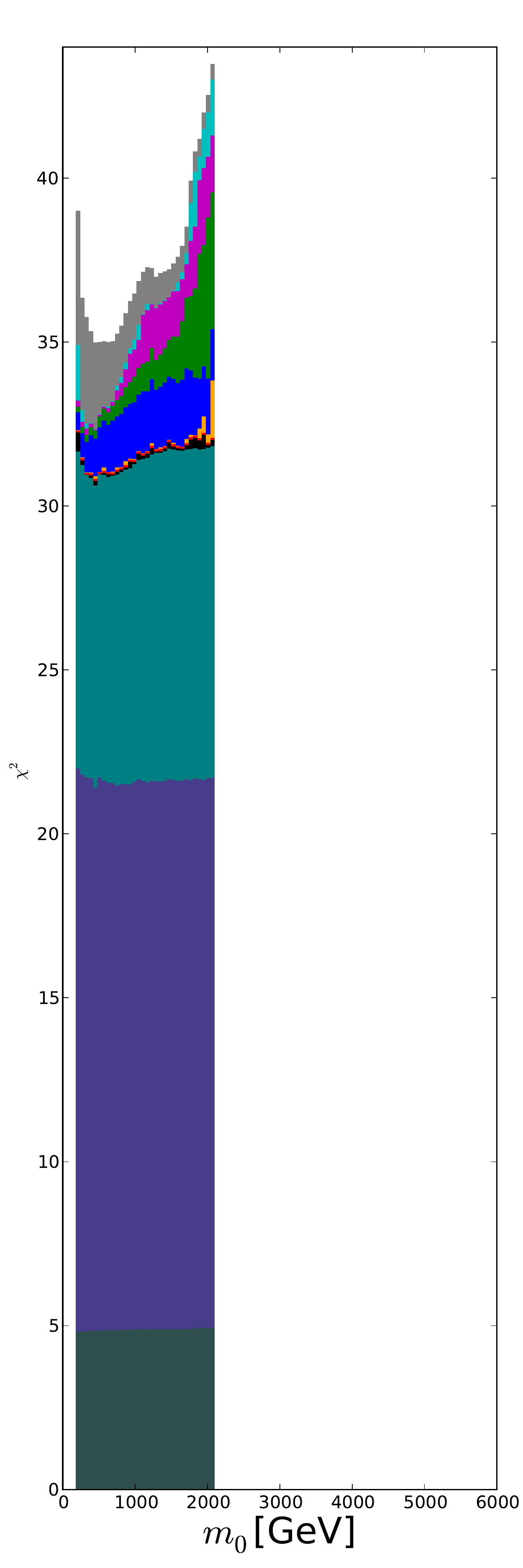}
\includegraphics[height=4in]{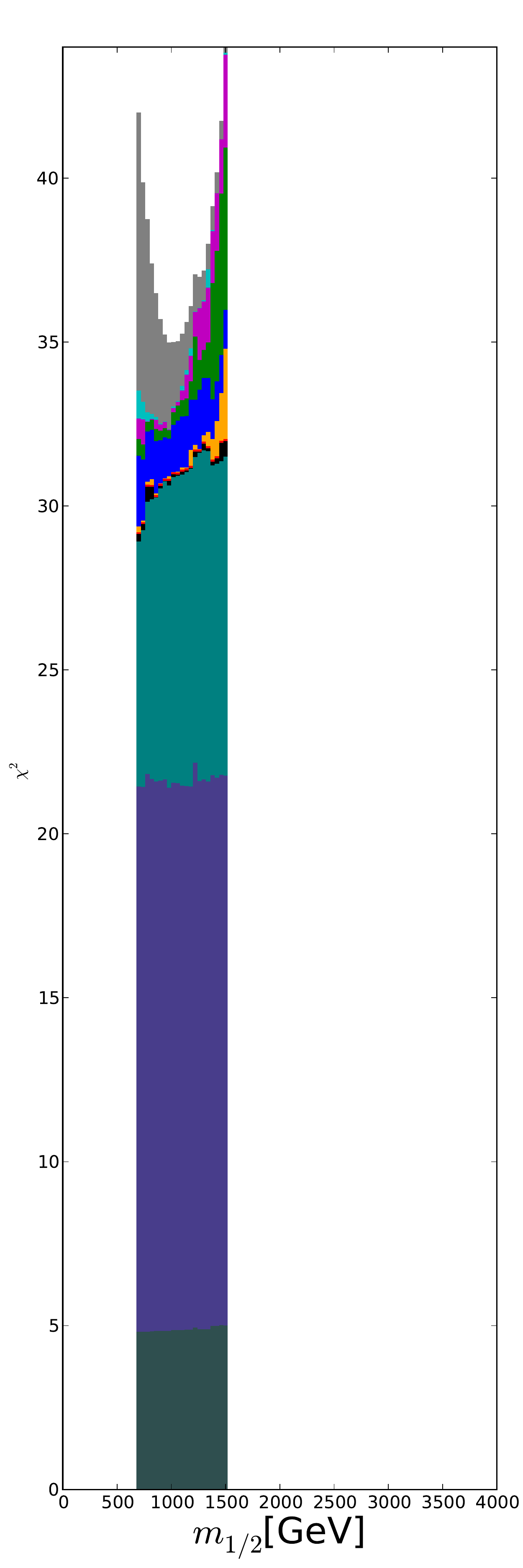}
\caption{\it Results from a recent global fit to the CMSSM showing (left column) the contributions to the global likelihood
at the best-fit point, and the marginalized likelihood as a function of (centre column) the $m_0$ mass and (right column)
$m_{1/2}$~\protect\cite{KdV}.}
\label{fig:CMSSMtowers}
\end{figure}

Fig.~\ref{fig:CMSSMtowers}~\cite{KdV} shows how the different observables contribute to building up the total
$\chi^2$ function in the CMSSM at the global minimum (left column), along the $m_0$ axis (central column), and
along the $m_{1/2}$ axis (right column). We see that the contribution of the flavour observables (dark grey) is
almost independent of $m_0$ and $m_{1/2}$, as is the contribution of the electroweak precision observables (purple).
On the other hand, the contribution from $g_\mu - 2$ (teal)~\cite{g-2} is quite large at the minimum and increases with both
$m_0$ and $m_{1/2}$. The contributions of the searches for the heavier supersymmetric Higgs bosons $H/A/H^\pm$,
and spin-independent dark matter scattering are never important, whereas the dark matter density constraint makes
itself felt at large $m_0$ and $m_{1/2}$. On the other hand, the $b \to s \gamma$ constraint loses importance at large
$m_0$ and $m_{1/2}$, whereas $B_s \to \mu^+ \mu^-$~\cite{CMSLHCb} gains in importance. The $W$ mass makes some contribution
at intermediate $m_0$ and $m_{1/2}$, whereas $m_H$ tends to disfavour both large and small $m_0$ and $m_{1/2}$.
Finally the LHC jets + MET constraint disfavours low values of $m_0$ and $m_{1/2}$, and the global mimumum
is largely determined by the tension between this constraint, $g_\mu - 2$ and, to lesser extents, $m_H$ and 
$B_s \to \mu^+ \mu^-$.

Fig.~\ref{fig:m0m12} also shows as solid (dashed) purple lines the present 95\% CL exclusion contour from searches for jets + MET
events during Run~1 of the LHC (prospective $5-\sigma$ CL jets + MET discovery sensitivity with 3000/fb of data at 14~TeV) for
the CMSSM, the NUHM1, the NUHM2 and the pMSSM10. We see that,
except in the case of the NUHM2, the global 95\% CL contour generally lies significantly beyond the current jets + MET
reach. This is due to the impacts of other constraints, notably the LHC measurement of the Higgs mass $m_H$, which, as was
seen in Fig.~\ref{fig:CMSSMtowers}, tends to favour relatively large sparticle masses. 

On the positive side, we see in Fig.~\ref{fig:m0m12} that significant portions of the 68\% CL regions in the CMSSM, the NUHM1 and the
NUHM2 lie within the prospective reach of the LHC with 3000/fb of data at 14~TeV, including in particular substantial
fractions of the regions where coannihilation with the stau is the dominant dark matter mechanism (shaded pink)~\cite{MC12}. In the case of the pMSSM10,
this mechanism does not play such an important r\^ole: the most important dark mechanism is coannihilation with charginos $\chi^\pm$
(shaded green)~\cite{MC12}. However, also in this case a significant portion of the 68\% CL region lies within reach of future LHC searches.

\section{A Long-Lived Charged Sparticle?}

\begin{figure*}[htb!]
\begin{center}
\resizebox{6.5cm}{!}{\includegraphics{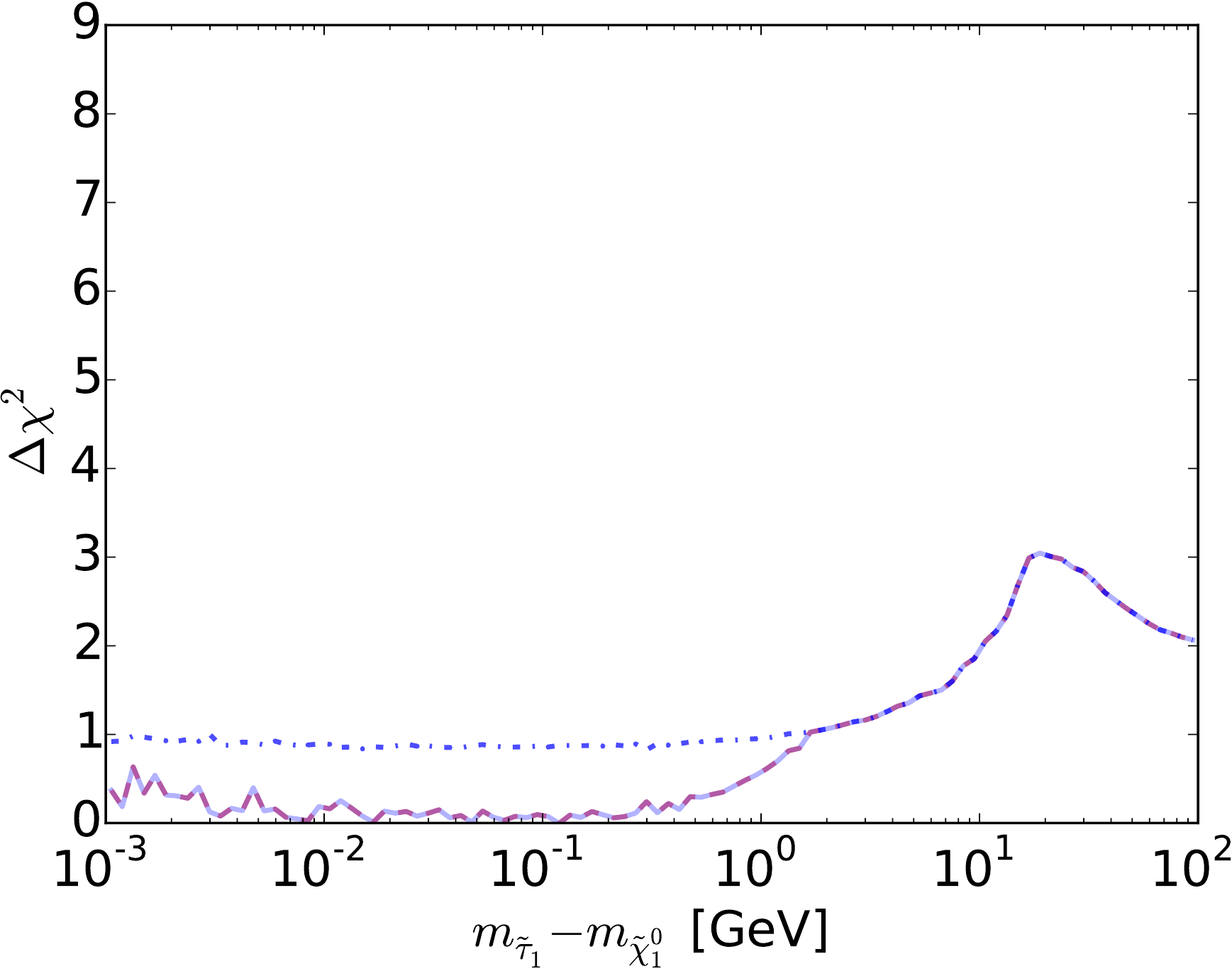}}
\hspace{0.8cm}
\resizebox{6.5cm}{!}{\includegraphics{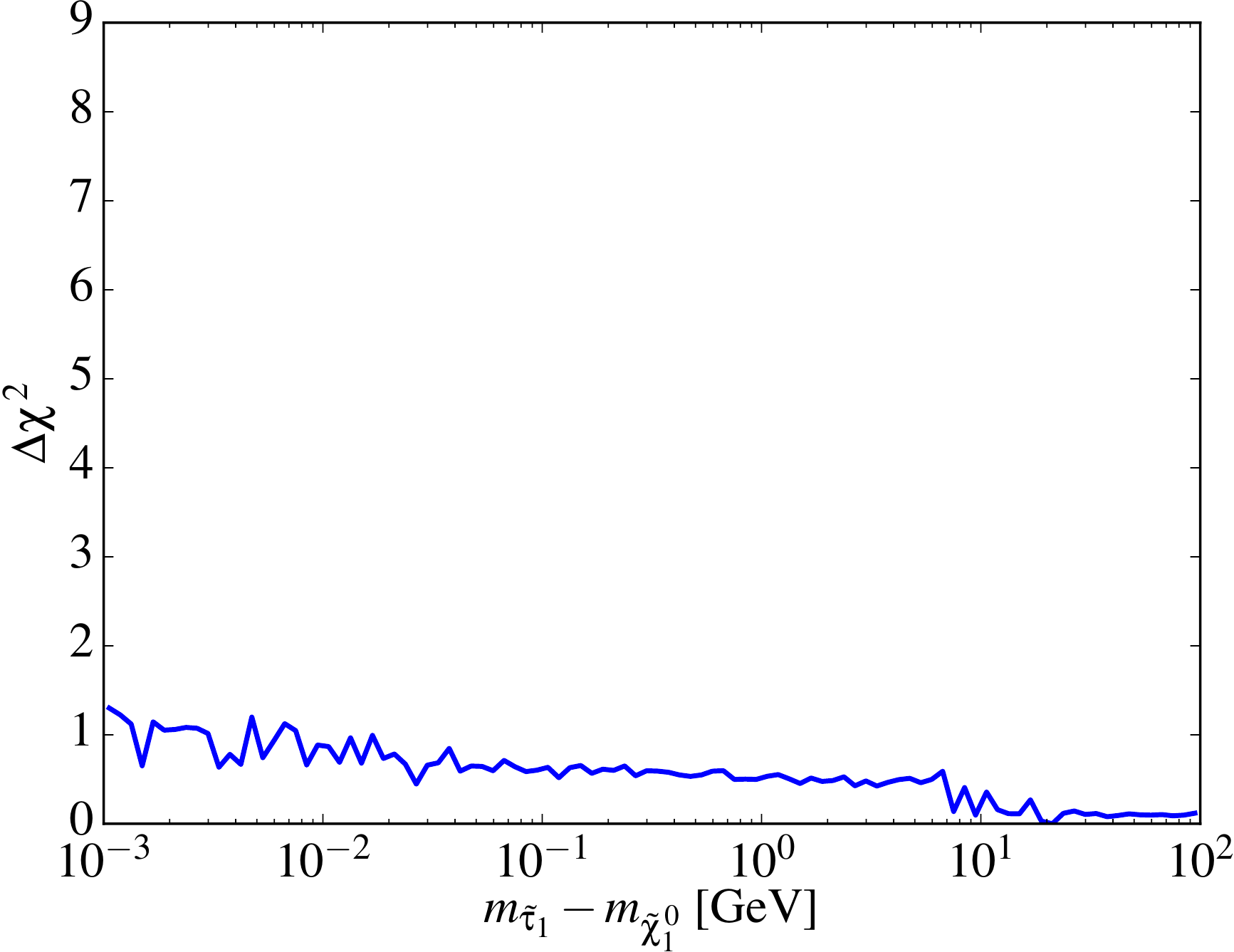}}\\
\resizebox{7.5cm}{!}{\includegraphics{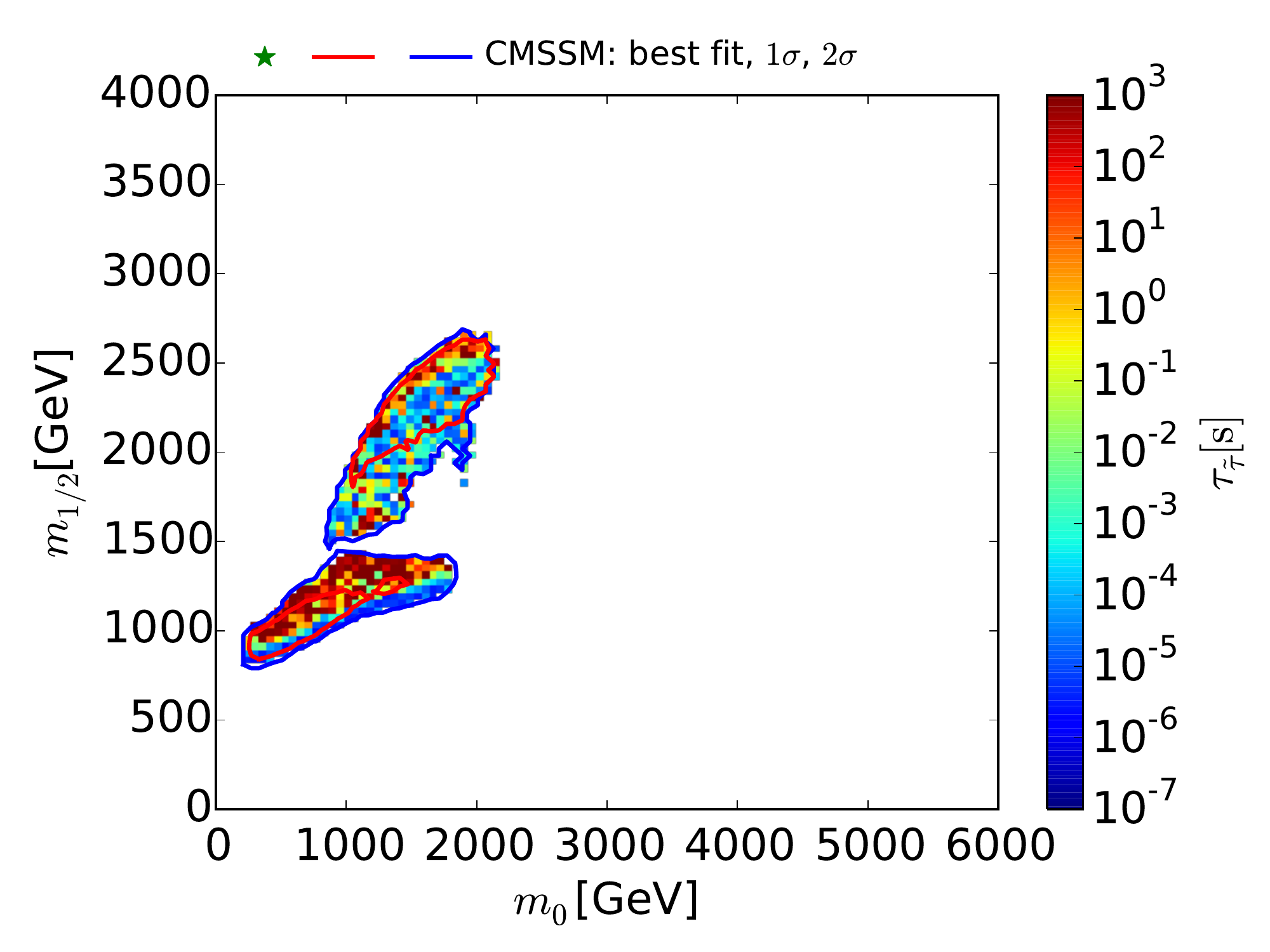}}
\resizebox{7.5cm}{!}{\includegraphics{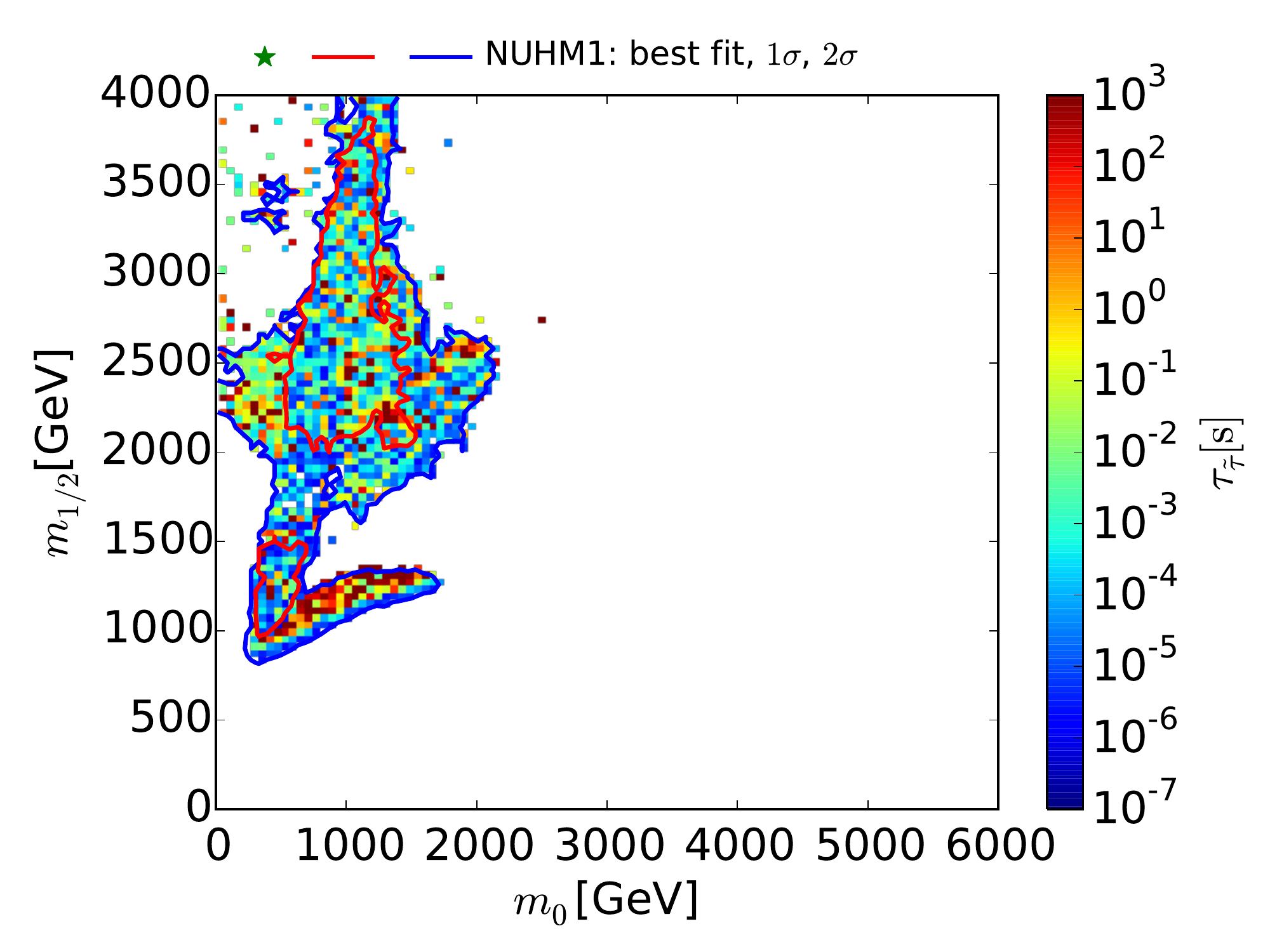}}\\
\end{center}
\vspace{-0.5cm}
\caption{\it Upper panels: the one-dimensional $\Delta \chi^2$ profile likelihood functions for $m_{\tilde \tau_1} - m_\chi$ in the CMSSM (left)
and the NUHM1 (right). Lower panels: the $(m_0, m_{1/2})$ planes in the CMSSM (left)
and the NUHM1 (right), with points colour-coded according to the ${\tilde \tau_1}$ lifetime found with the best-fit parameters
for the corresponding values of $(m_0, m_{1/2})$, and displaying only points with lifetimes $\in [10^{-7}, 10^3]$~s~\protect\cite{MC12}.}
\label{fig:stauchi}
\end{figure*}

What would be the Run-2 signatures of supersymmetry in in these model scenarios? Clearly jets + MET will continue to
be key, but other signatures may also be important. In particular, in the cases of the CMSSM, the NUHM1 and the NUHM2, there are
substantial possibilities for detecting a long-live charged supersymmetric particle. In the pink-shaded regions of Fig.~\ref{fig:m0m12}
the next-to-lightest supersymmetric particle (NLSP) is the lighter stau and the dark matter density is brought into the
cosmological range by LSP-stau coannihilation. In this case the stau-LSP mass difference may be very small, as
illustrated in the upper panels of Fig.~\ref{fig:stauchi} for the CMSSM (left panel) and the NUHM1 (right panel)~\cite{MC12}.
The mass difference may be 1~GeV or (much) less with a $\Delta \chi^2$ penalty $\lesssim 1$. This can lead to a
relatively long lifetime for the ${\tilde \tau_1}$, as seen in the lower panels of Fig.~\ref{fig:stauchi}, where the
distributions in the $(m_0, m_{1/2})$ planes of the CMSSM (left panel) and the NUHM1 (right panel) of model
parameter sets with ${\tilde \tau_1}$ lifetimes in the range $[10^{-7}, 10^3]$~s are shown~\footnote{ If the stau had a
longer lifetime, its decays would modify the successful predictions of conventional Big-Bang nucleosynthesis.}. As in Fig.~\ref{fig:m0m12},
the red and blue contours delineate the $\delta \chi^2 = 2.30$ and $5.99$ regions of these long-lived stau samples.

Charged sparticles with lifetimes $\gtrsim 10^{-7}$~s are likely to exit an LHC detector before decaying,
providing metastable charged particle signatures such as a non-relativistic time of flight and anomalously high ionization.
On the other hand, if the stau decays inside the detector, it may provide a disappearing-track signature.
Supersymmetric final states at the LHC originate mainly from cascade decays of heavier sparticles. Therefore
they may contain zero, one or two long-lived staus, and hence produce combinations
of these signatures and also jets + MET~\cite{CELMOV}. A study has shown that the full extent of the stau coannihilation strip
of the CMSSM can be explored by searching for these signatures at Run~2 of the LHC~\cite{DELM}. So far, the LHC
experiments have analyzed their searches for long-lived sparticles mainly in the framework of near-degeneracy
between the lightest chargino and neutralino: it is to be hoped that in the future these analyses will be extended to 
the case of a long-lived stau.

\section{Global Analysis of the pMSSM10}

\begin{figure}[htb!]
\resizebox{7.5cm}{!}{\includegraphics{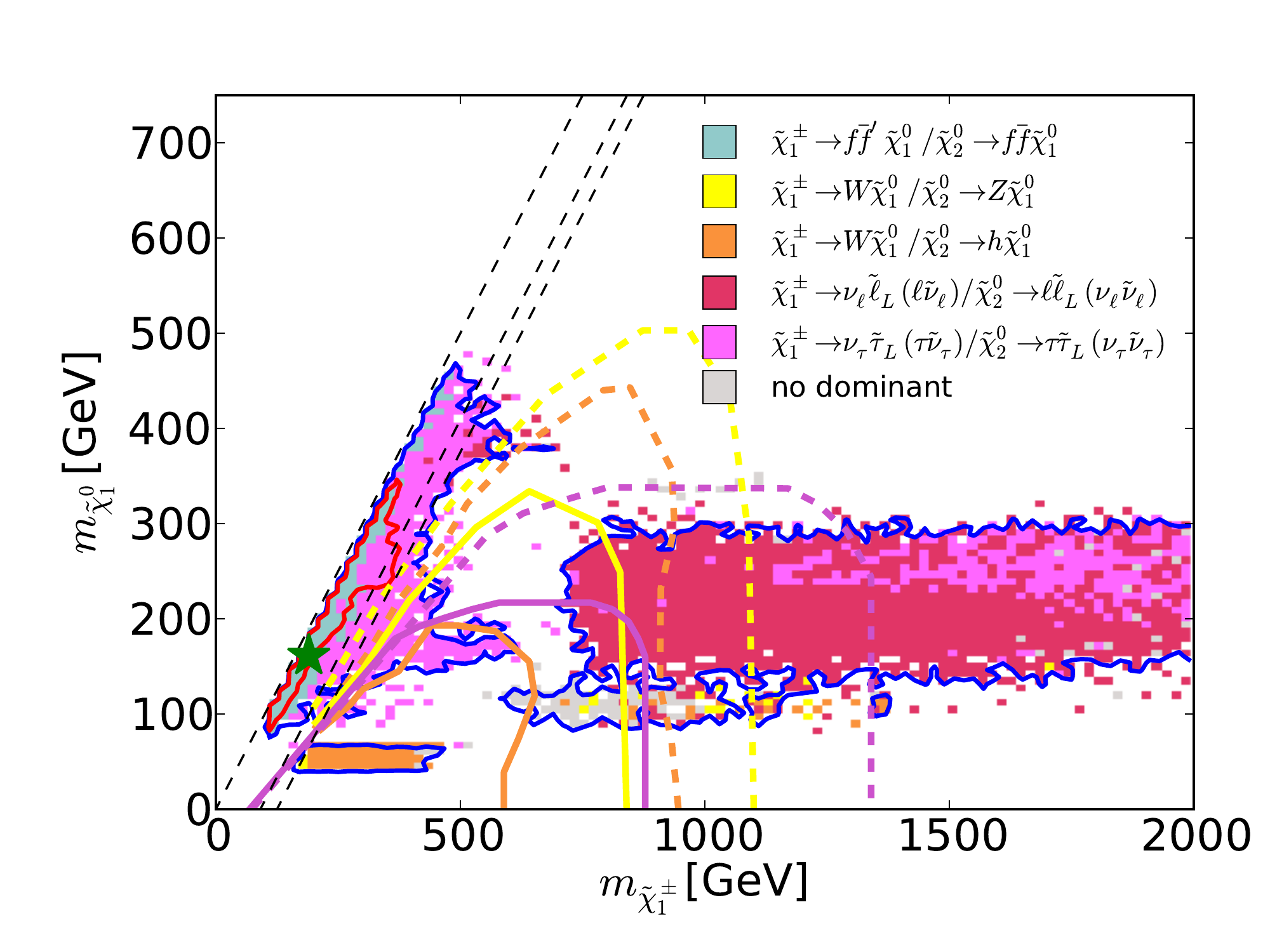}} 
\resizebox{7.5cm}{!}{\includegraphics{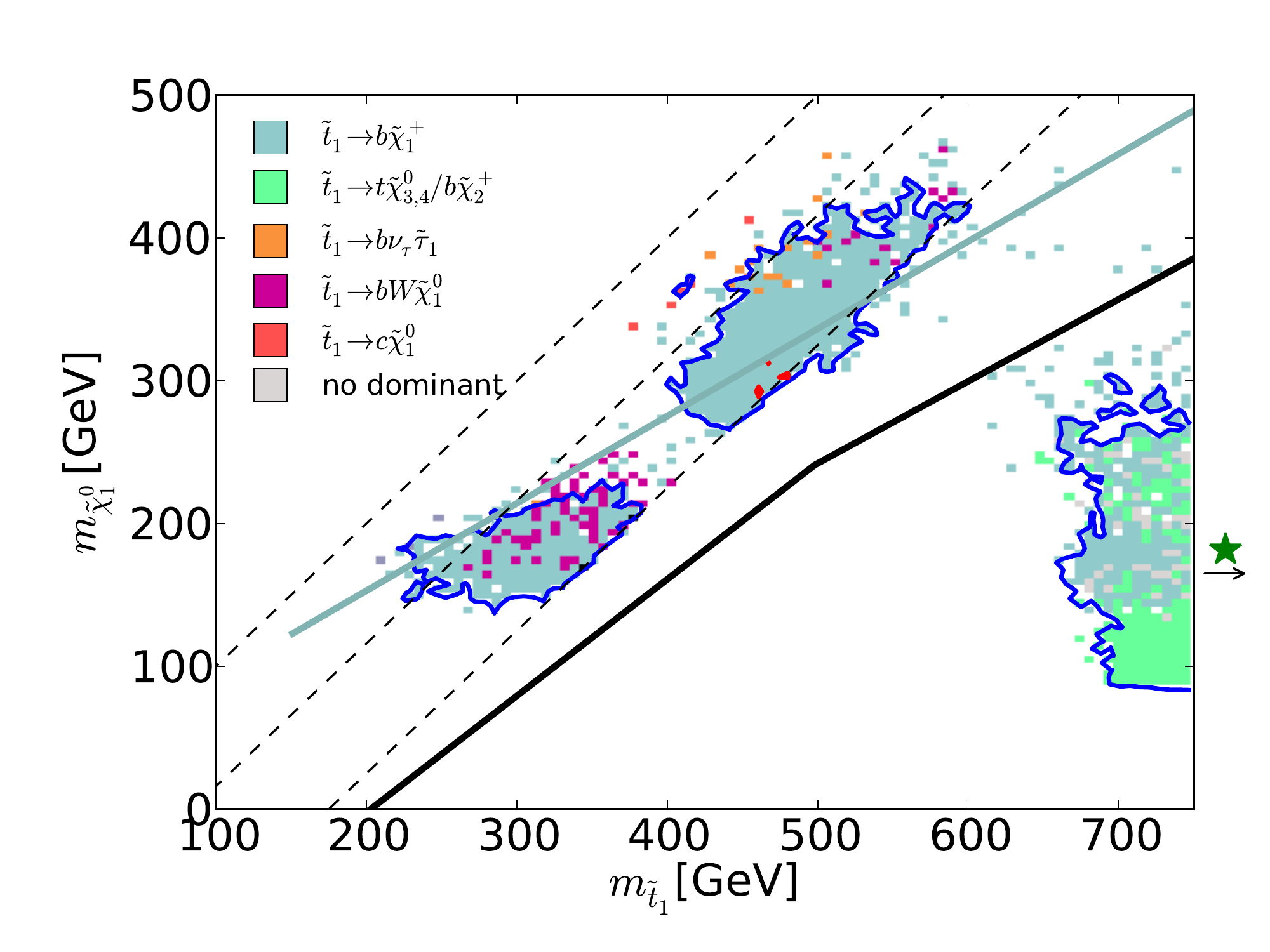}}
\caption{\it Upper left panel: The $(m_{\chi^\pm_1}, m_\chi)$ plane with our 
68 and 95\% CL contours shown as solid red and blue lines, respectively.
The coloured shadings indicate where the corresponding branching ratios exceed 50\%.
Also shown as solid (dashed) yellow/orange/purple lines are the
projected LHC 95\% CL$_s$ exclusion reaches
for associated $\chi^\pm_1$ and $\chi_2$ production with decays via 
$W/Z$/$W/h$/${\tilde \ell_L}/{\tilde \nu_{\ell_L}}$/${\tilde \tau_L}/{\tilde \nu_{\tau_L}}$ with 300 (3000)/fb of data, if these decays are dominant. 
Upper right panel: The $(m_{\tilde t_1}, m_\chi)$ plane with our 
68 and 95\% CL contours shown as solid red and blue lines, respectively, 
as well as coloured regions where the indicated branching ratios
exceed 50\%. The projected LHC sensitivity with 300/fb for ${\tilde t_1} \to \chi + t$ decays
is shown as a thick black line, 
and the corresponding
sensitivity for ${\tilde t_1} \to \chi^\pm_1 b$ decays (if they are dominant) is shown as a pale blue dashed line~\protect\cite{MC11}.}
\vspace{1em}
\label{fig:pMSSM10-planes}
\end{figure}

In contrast to the CMSSM, the NUHM1 and the NUHM2, in the pMSSM10~\cite{MC11}  the NLSP is usually the
lighter chargino ${\chi^\pm_1}$, but the mass difference between the LSP and the NLSP
is not expected to be very small, and a long-lived charged sparticle is disfavoured.
Many searches with the LHC at 8~TeV bear upon the allowed regions of the pMSSM10
parameter space, which were analysed using the {\tt Fastlim/Atom}~\cite{Atom} and {\tt Scorpion}~\cite{Scorpion} codes.
The left panel of Fig.~\ref{fig:pMSSM10-planes} displays the $(m_{\chi^\pm_1}, m_\chi)$ plane
in the pMSSM10, with the 68\% and 95\% CL regions outlined as usual in red and blue, respectively.
The diagonal dashed black lines correspond to $\Delta m \equiv m_{\tilde \chi^\pm_1} - m_\chi = 0, M_Z$ and $m_H$. 
The coloured shadings correspond to the dominant ${\chi^\pm_1}$ decay modes, and the orange, yellow
and purple solid (dashed) lines represent the estimated reaches with 300 (3000)/fb of LHC data,
assuming 100\% branching ratios.

The pMSSM10 also offers the possibility of a relatively light stop squark, as seen in the right panel
of Fig.~\ref{fig:pMSSM10-planes}. Here again, the many LHC searches that constrain the allowed regions of the pMSSM10
parameter space were analysed using the {\tt Fastlim/Atom}~\cite{Atom} and {\tt Scorpion}~\cite{Scorpion} codes.
The diagonal dashed lines are where $\Delta m \equiv m_{\tilde t_1} - m_\chi = 0, M_W + m_b$
and $m_t$. The light blue shading shows that ${\tilde t_1} \to b {\tilde \chi^\pm_1}$ is the dominant decay
in much of the region displayed, and the sensitivity to this decay of the LHC with 300/fb is shown as a solid line in
the corresponding colour (the reach with 3000/fb is similar). The solid black line shows the projected reach for ${\tilde t_1} \to t \chi$ if this is
the dominant decay, which is however not the case in our pMSSM10 analysis.

\begin{figure}[htb]
\centering
\includegraphics[height=4in]{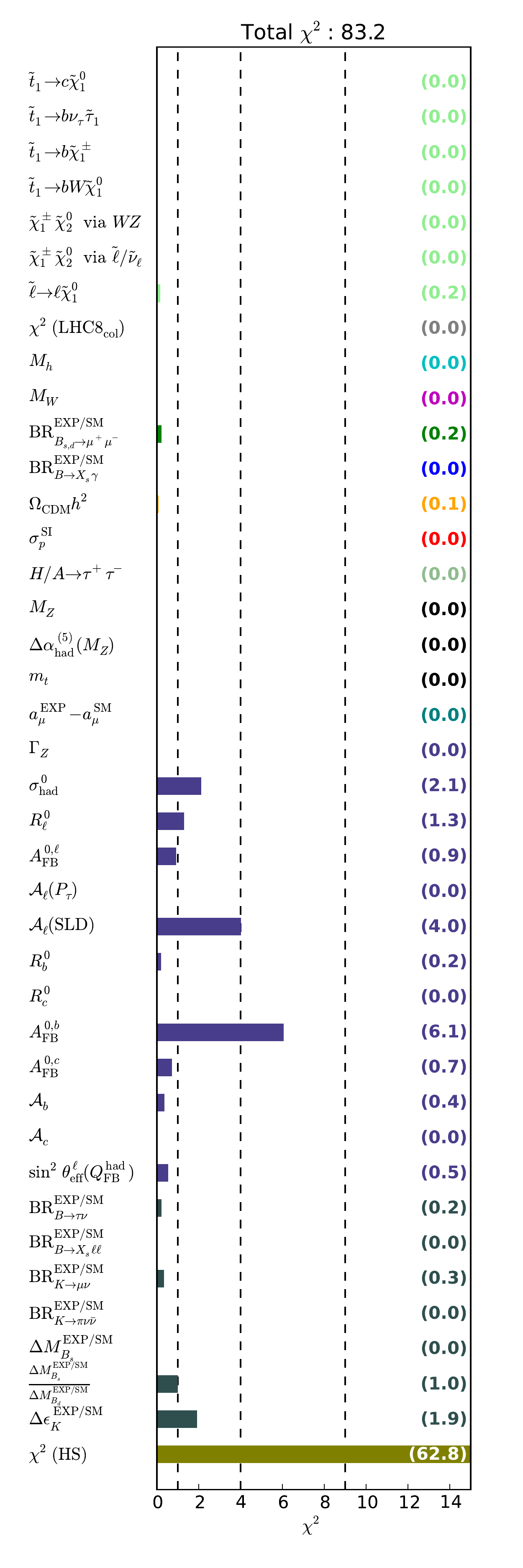}
\includegraphics[height=4in]{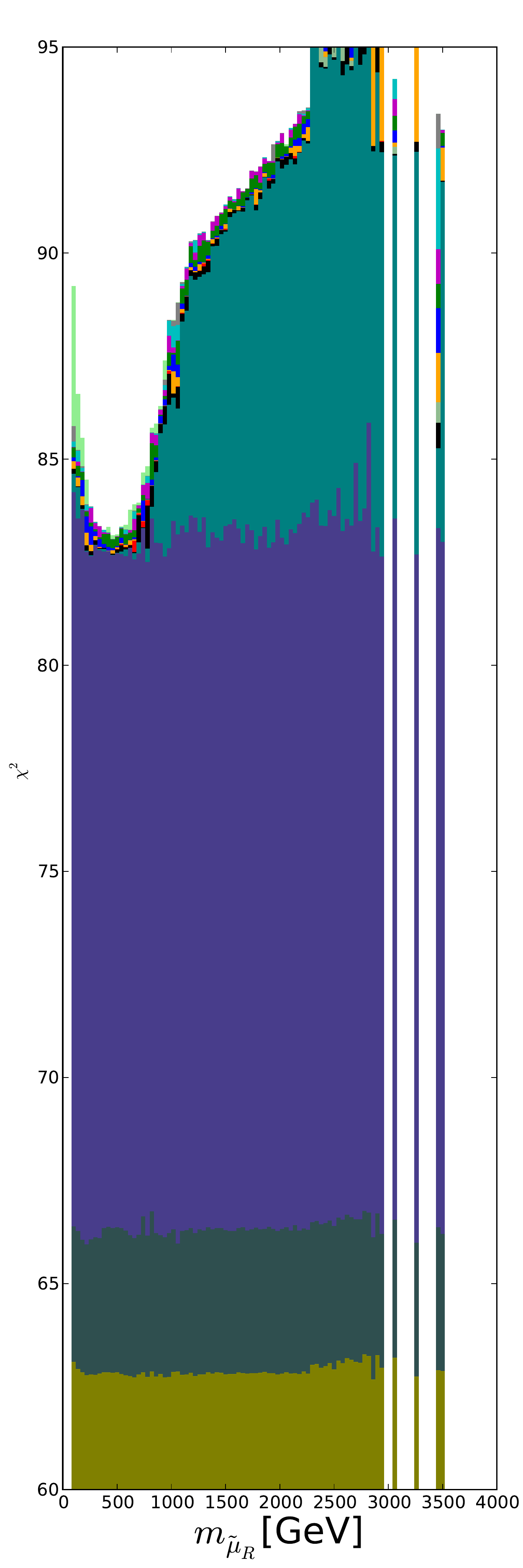}
\includegraphics[height=4in]{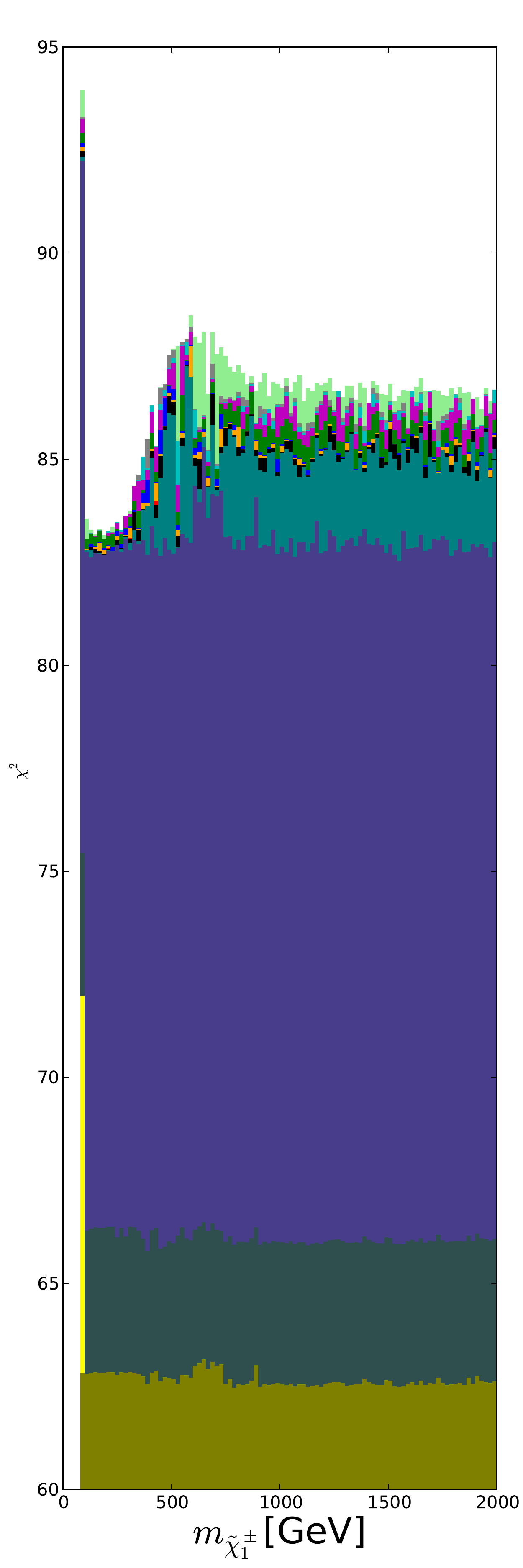}
\caption{\it Results from a recent global fit to the pMSSM10 showing (left panel) the contributions to the global likelihood
at the best-fit point, and the marginalized likelihood as a function of (centre panel) the ${\tilde \mu_R}$ mass and (right panel)
the lighter chargino mass, $m_{\tilde \chi^\pm_1}$~\protect\cite{KdV}.}
\label{fig:pMSSM10towers}
\end{figure}

Fig.~\ref{fig:pMSSM10towers} displays the breakdown of the global $\chi^2$ function in our analysis of the pMSSM10>
We see that the Higgs production and decay data (olive green analysed using the {\tt HiggsSignals}~\cite{HiggsSignals} code,
notice the suppressed zero), the flavour
observables (green) and the precision electroweak measurements (purple) make contributions that are rather insensitive
to the values of the lighter smuon mass $m_{\tilde \mu_R}$ (centre column) and the LSP mass (right column).
We also find relatively unimportant contributions from the dark matter constraints, $B_s \to \mu^+ \mu^-$
and the LHC searches. The shape of the global $\chi^2$ function is mainly determined by $g_\mu - 2$ (teal),
which exerts a strong preference for small $m_{\tilde \mu_R}$ and also favours small $m_\chi$.

\section{{\it Floreat} $g_\mu - 2$?}

\begin{figure*}[htb!]
\begin{center}
\resizebox{9cm}{!}{\includegraphics{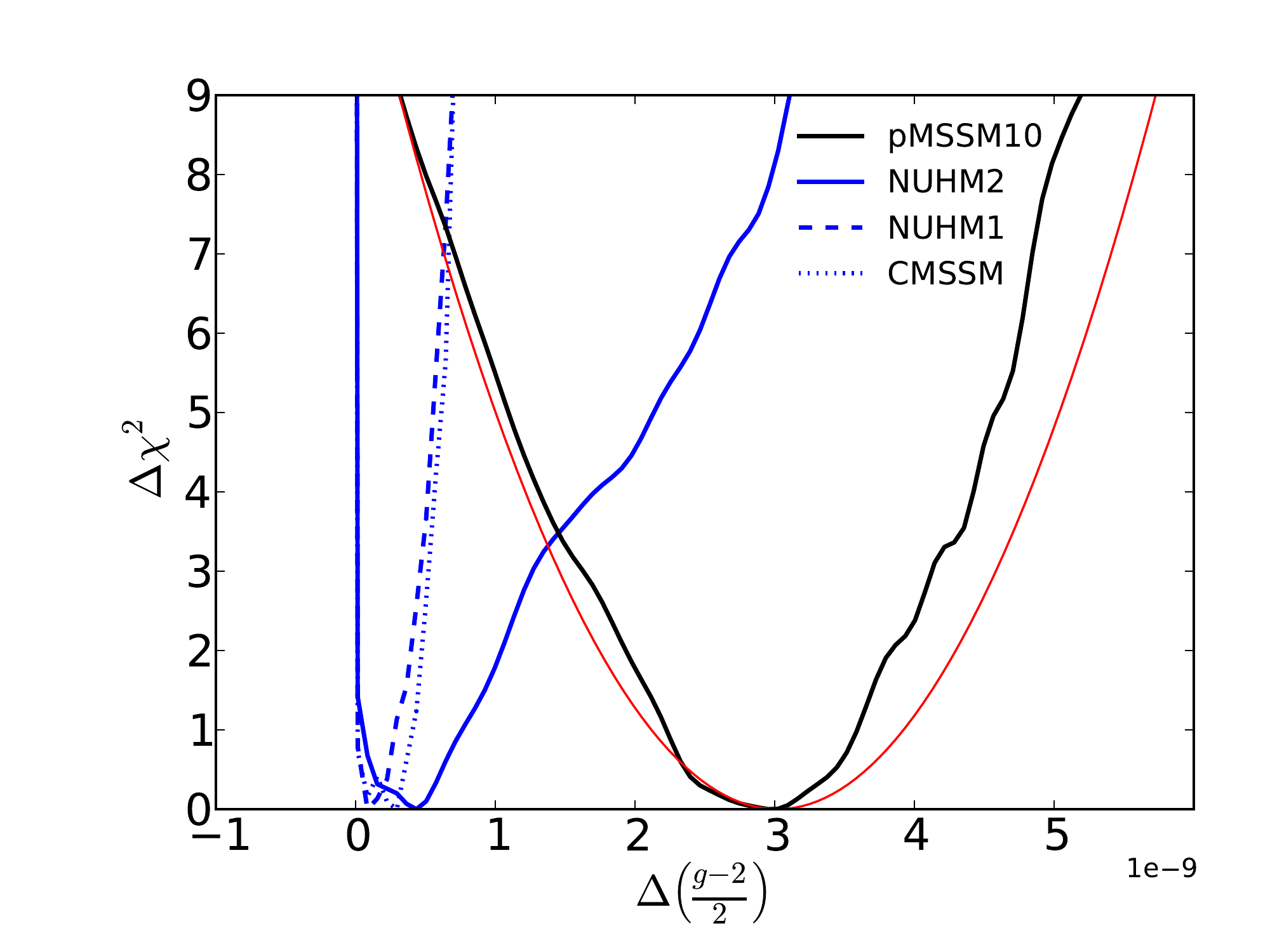}}
\end{center}
\vspace{-0.5cm}
\caption{\it One-dimensional profile likelihoods for the supersymmetric contributions to $g_\mu - 2$, from recent global
fits to the  CMSSM (blue dotted line), the NUHM1 (blue dashed line), the NUHM2 (blue solid line) and
the pMSSM10 (black solid line), with the experimental likelihood (solid red line) shown for comparison~\protect\cite{MC11}. 
}
\label{fig:gmt}
\end{figure*}

This preference for small $m_{\tilde \mu_R}$ reflects the fact that, in contrast to the CMSSM and
the NUHM1,2, the pMSSM10 can reconcile $g_\mu - 2$ with the absence of supersymmetry at the
LHC (so far)~\cite{MC11}. This contrast is also visible in Fig.~\ref{fig:gmt}, where we see that the pMSSM10
(solid black curve) can fit perfectly the experimental measurement of $g_\mu - 2$ (solid red curve), whereas
the CMSSM, the NUHM1 and the NUHM2 (blue solid, dashed and dotted lines, respectively) prefer
values of $g_\mu - 2$ that are close to the Standard Model value. In our analysis, this differs from
the experimental measurement by $\sim 3 \sigma$. It is this inability to fit $g_\mu - 2$ that is
largely responsible for the poor overall quality of the the fits in the CMSSM, the NUHM1 and the NUHM2 (and also the SM).
Naive estimates using the $\chi^2$ function indicate $p$-values below 10\% for these models,
and a similar estimate has been obtained in~\cite{Fittino} using toys. On the other hand, we estimate
a $p$-value $\sim 30$\% for the pMSSM10.

This comparison shows that elucidating the apparent discrepancy between the experimental measurement and the
Standard Model calculation of $g_\mu - 2$ is crucial for understanding the prospects for discovering supersymmetry
in the near future. Is it possible that the theoretical calculation suffers from an unknown systematic uncertainty,
possibly in the calculation of the light-by-light scattering contribution? Or is there some experimental effect that
remains to be understood? If not, some new physics at the TeV scale is needed, and supersymmetry fits the bill,
particularly within the pMSSM framework.
The good news is that an experiment to remeasure $g_\mu - 2$ with significantly improved accuracy is in an
advanced stage of preparation at FNAL~\cite{FNALg-2}.

\section{Prospects for Discovering Supersymmetry at the LHC}

\begin{figure*}[htb!]
\resizebox{7.5cm}{!}{\includegraphics{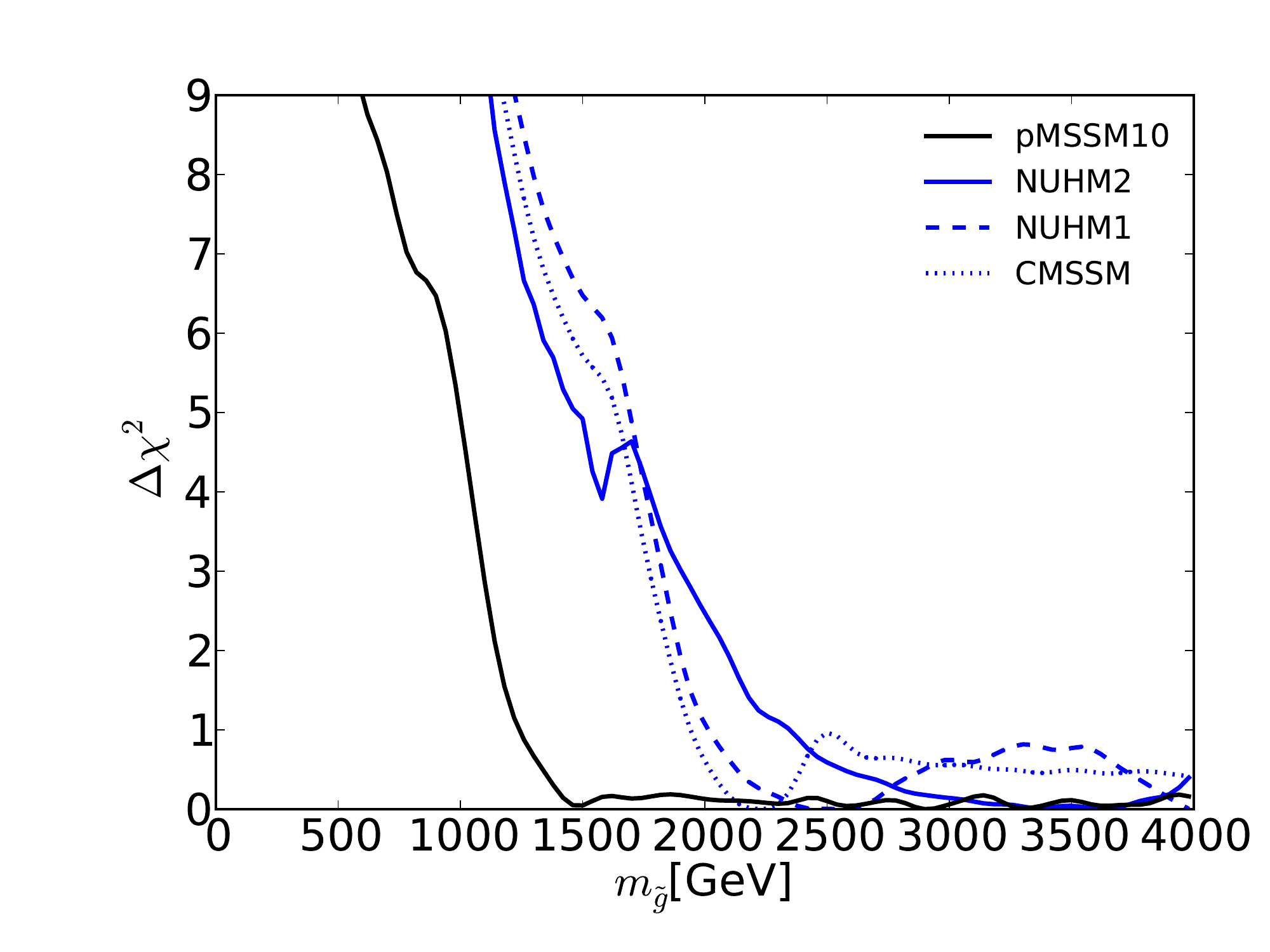}}
\resizebox{7.5cm}{!}{\includegraphics{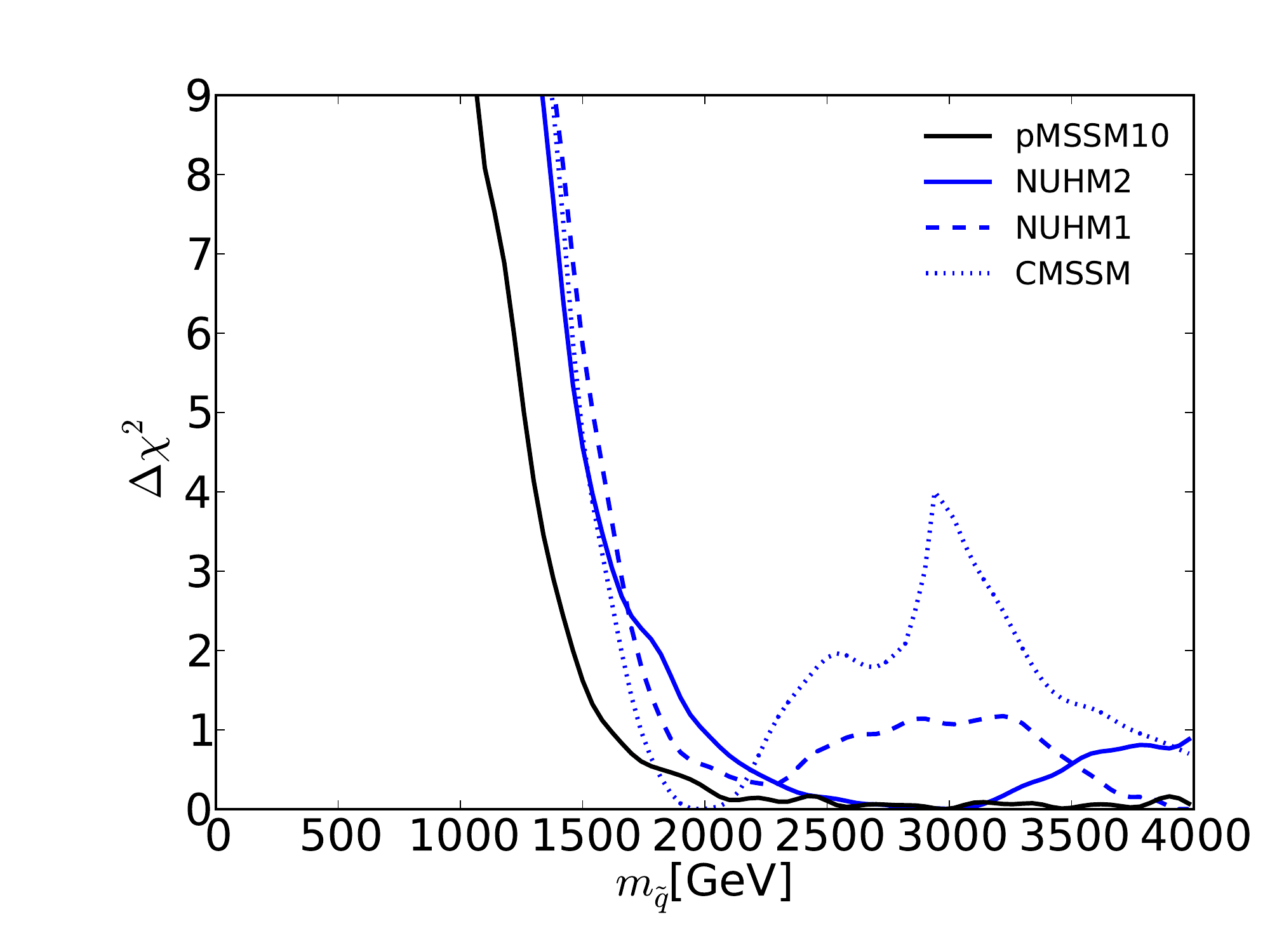}}\\
\vspace{-0.5cm}
\caption{\it One-dimensional profile likelihood functions
for $m_{\tilde g}$ and $m_{\tilde q}$ from recent global analyses of MSSM scenarios. 
In each panel the solid black line is for the pMSSM10, the solid blue line for the NUHM2,
the dashed blue line for the NUHM1, and the dotted blue line for the
CMSSM~\protect\cite{MC11}.
}
\label{fig:mgmsq}
\end{figure*}

Fig.~\ref{fig:mgmsq} displays the marginalised global $\chi^2$ functions for the masses of the gluino (left panel)
and the squark (right panel) in the CMSSM (solid blue lines), the NUHM1 (dashed blue lines), the NUHM2 (dotted blue lines)
and the pMSSM10 (solid black line)~\cite{MC11}. In the case of the gluino, we see that the pMSSM10 offers better prospects
than the other models for a relatively light gluino with mass $\sim 1500$~GeV, whereas the model predictions are
more similar for the squark mass. Here the good news is that jets + MET searches at the LHC will become sensitive
to $m_{\tilde g}, m_{\tilde q} \sim 3$~TeV with 3000/fb of integrated luminosity at 14~TeV, so there are decent
prospects (but no guarantees) for discovering supersymmetry at the LHC in this channel.

As we have seen previously, in the CMSSM, the NUHM1 and the NUHM2 there are also interesting prospects
for discovering long-lived charged sparticles at the LHC, if the DM density is brought into the cosmological
range by stau coannihilation. This is not likely in the pMSSM10, but in this model searches for charginos
may be promising, as discussed in the previous Section. Table~\ref{table:modelsensitivity}~\cite{MC12} compiles the
prospects for supersymmetry searches at the LHC in the CMSSM, the NUHM1, the NUHM2 and the pMSSM10,
organised according to the dominant mechanisms for bringing the relic LSP density into the range allowed
by cosmology. Also shown are the prospects for direct DM detection in each case. We see that in every
instance where some mechanism may become dominant, there are prospects for detection either at the LHC or in
direct DM search experiments (or both).

\begin{table}[h!]
	\center
	{\small
	\begin{tabular}{ | c || c || c | c | c | c|}
		\hline
		DM & Exp't & \multicolumn{4}{c|}{Models} \\ 
		mechanism & & CMSSM & NUHM1 & NUHM2 & pMSSM10 \\ \hline
		${\tilde \tau_1}$ & LHC & {$\checkmark$ MET, $\checkmark$ LL} & ($\checkmark$ MET, $\checkmark$ LL) & 
		($\checkmark$ MET, $\checkmark$ LL) & ($\checkmark$ MET), $\times$ LL \\ 
		coann. & DM & ($\checkmark$) & ($\checkmark$) & $\times$ & $\times$ \\ \hline
		$\chi^\pm_1$ & LHC & - & $\times$ &  $\times$ & ($\checkmark$ MET)  \\ 
		coann. & DM & - & $\checkmark$ & $\checkmark$ & ($\checkmark$)  \\ \hline
		${\tilde t_1}$ & LHC & - & - & $\checkmark$ MET & -  \\ 
		coann. & DM & - & - & $\checkmark$ & -  \\ \hline
		$A/H$ & LHC & $\checkmark$ $A/H$ & ($\checkmark$ $A/H$) & ($\checkmark$ $A/H$) & -  \\
		funnel & DM & $\checkmark$ & $\checkmark$ & ($\checkmark$) & - \\ \hline
		Focus & LHC & ($\checkmark$ MET) & - & - & -  \\
		point & DM & $\checkmark$ & - & - & - \\ \hline
		$h,Z$ & LHC & - & - & - & ($\checkmark$ MET) \\
		funnels & DM & - & - & - & ($\checkmark$) \\ \hline
			\end{tabular}}
			\vspace{0.5cm}
	\caption{\it Compilation of assessments of the detectability of supersymmetry in the CMSSM, NUHM1, NUHM2 and pMSSM10
	models at the LHC in searches for MET events, long-lived particles charged (LL) and heavy $A/H$
	Higgs bosons, and in direct DM search experiments, depending on the dominant mechanism for
	bringing the DM density into the cosmological range. The symbols $\checkmark$, ($\checkmark$) and
	$\times$ indicate good prospects, interesting possibilities and unlikely prospects, respectively.
	The symbol - indicates that a DM mechanism is not important for the corresponding model~\protect\cite{MC12}.}
	\label{table:modelsensitivity}
\end{table}

\section{Prospects for Measuring Supersymmetric Model Parameters}

Given the interesting prospects for discovering supersymmetric particles during future LHC runs,
what are the prospects for measuring supersymmetric model parameters? The answer to this
question is highly model-dependent, and the experiments would surely do much better than we
theorists can currently imagine.

\begin{figure}[ht!]
\centerline{
\includegraphics[height=5.5cm]{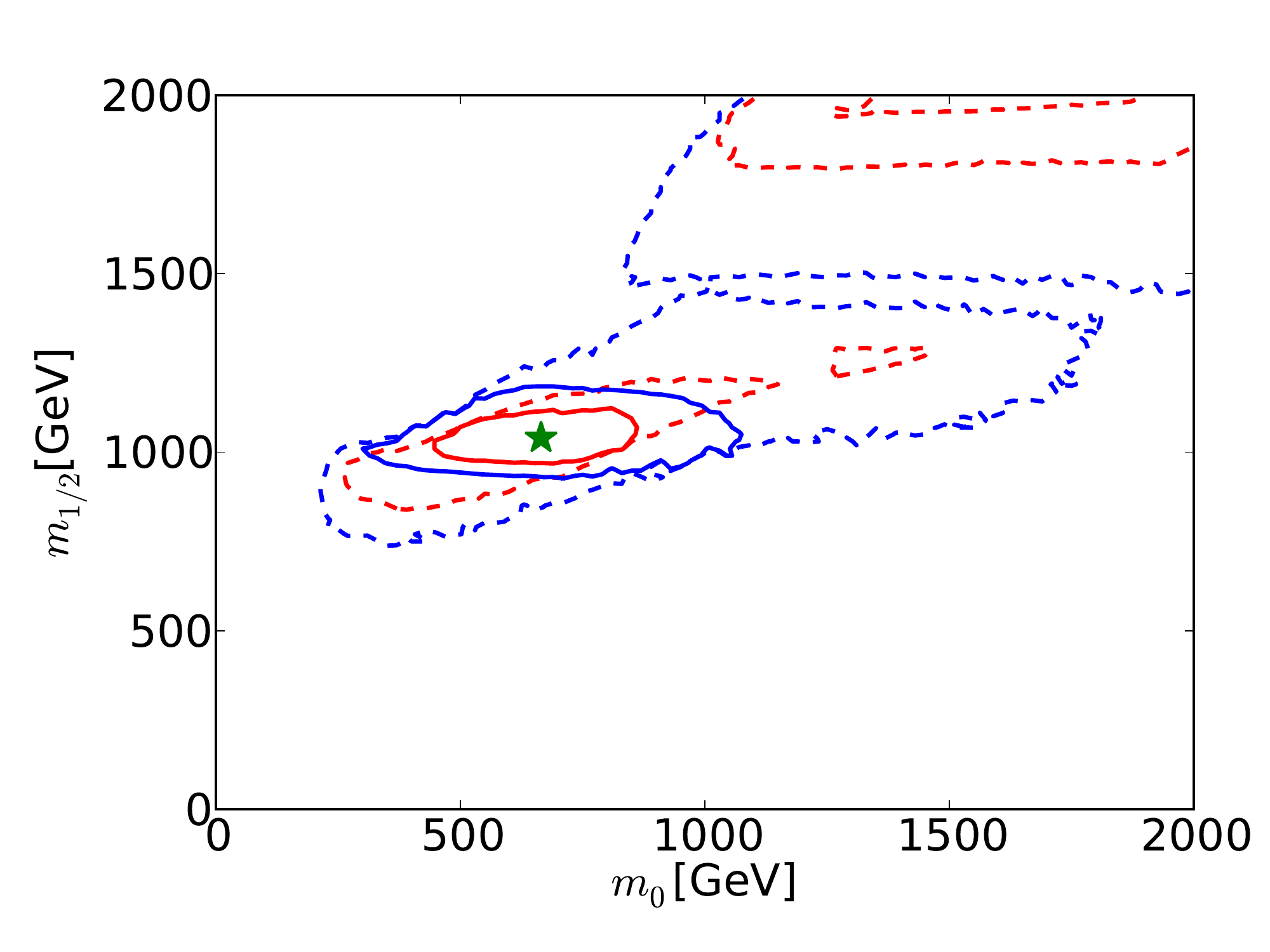}
\includegraphics[height=5.5cm]{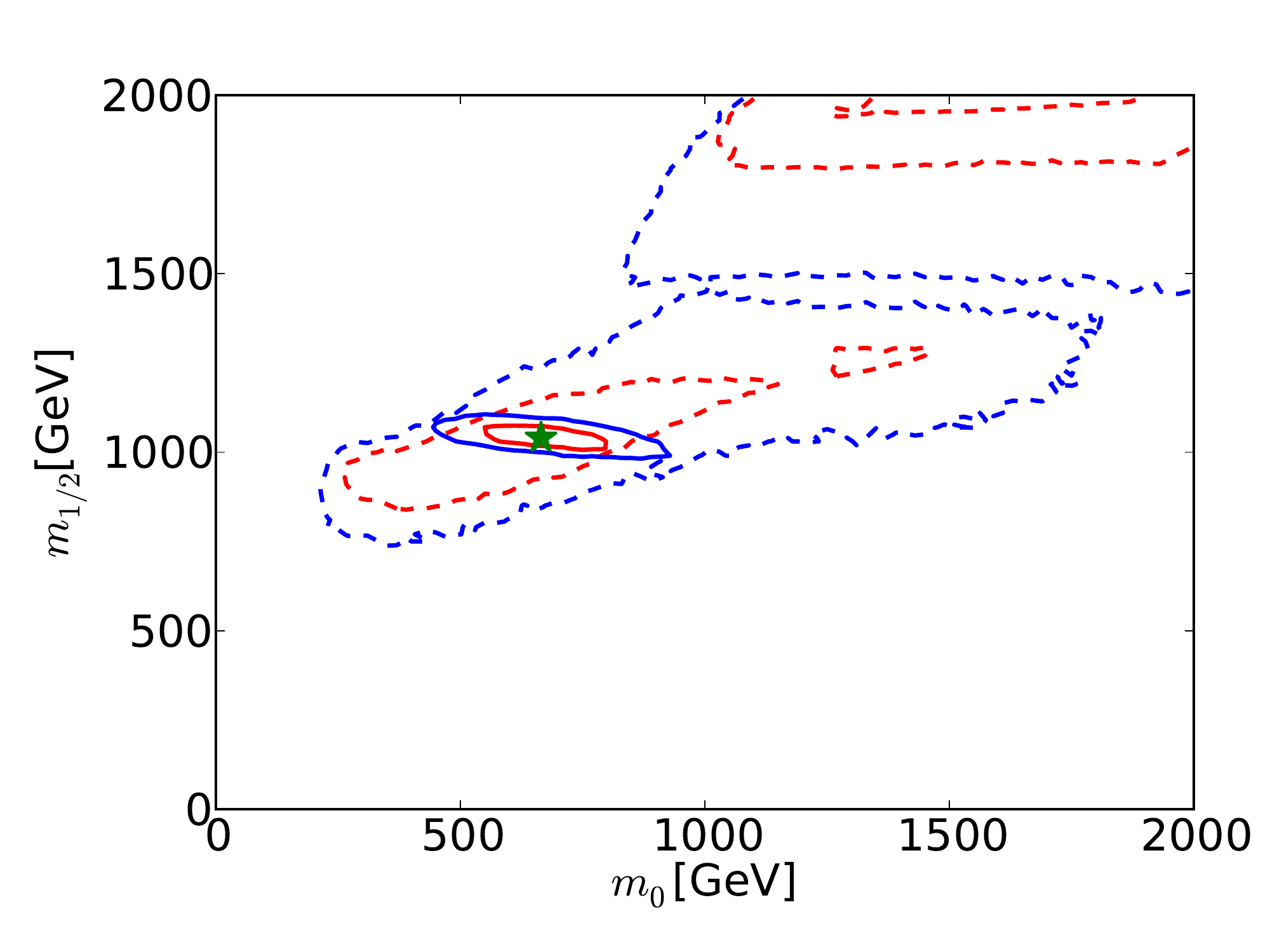}
}
\caption{\label{fig:m0m12Jad}\it
The solid lines show the prospective results of fits combining LHC measurements of the cross-section, MT2 and supplementary jets
at LHC14 with 300/fb (left panel) and 3000/fb (right panel) with the results of a recent global fit to the CMSSM (dashed lines): the
red and blue contours represent the 68 and 95\% CL regions, respectively~\protect\cite{Interplay}.}
\end{figure}

Fig.~\ref{fig:m0m12Jad}~\cite{Interplay} shows some results from an exploratory study the simplest case of the CMSSM and the
impact of the entry-level jet measurements of events containing squarks and gluinos. This analysis assumed that Nature
has chosen the best-fit point found in the above-mentioned global analysis of the CMSSM, namely $m_0 = 670$~GeV, 
$m_{1/2} = 1040$~GeV, $A_0 = 3440$~GeV and $\tan \beta = 21$. Included in this
analysis were prospective measurements of the total cross section for jet + MET events, the distribution in the
MT2 variable, and the spectator jet energies in ${\tilde g} \to {\tilde q_R} + {\bar q}$ decay. Generally speaking,
these measurements are more sensitive to the value of $m_{1/2}$ than to $m_0$, as seen in the left panel of
Fig.~\ref{fig:m0m12Jad} that assumes 300/fb of integrated luminosity at 14~TeV, and the right panel that
assumes 3000/fb of integrated luminosity. The dashed red (blue) lines bound the 68 (95)\% CL regions found
in the previous global analysis, and the solid red (blue) lines show the result on combining the prospective LHC
measurements with this global analysis. We see that measurements with 300/fb would already restrict quite strongly
the preferred region of parameter space, and that the effects of measurements with 3000/fb would be extremely
restrictive.

This analysis shows that the LHC has good prospects for confronting indirect estimates of
supersymmetric model parameters with more direct measurements. A successful confrontation,
as shown in the CMSSM case in Fig.~\ref{fig:m0m12Jad}, would provide a non-trivial verification of 
supersymmetry at the quantum level. This could parallel the previous tests of the SM at the quantum level
that led to successful predictions of the top and Higgs masses.

\section{Prospects for Discovering Supersymmetry at Future Colliders}

What if supersymmetry does not turn up at the LHC? What are the prospects for discovering it at
some future collider? The constraints from the LHC already disfavour, but do not exclude, the possibility
of discovering supersymmetry at a future $e^+ e^-$ collider. There is not much scope for discovering it
at the ILC with its centre-of-mass energy $\le 500$ or 1000~GeV, and the same is true {\it a fortiori}
for the CEPC or FCC-ee with their more limited centre-of-mass energies. These machines would
probably be limited to indirect probes of supersymmetry, at which FCC-ee would excel~\cite{Interplay}. However, CLIC
with its higher centre-of-mass energy $\le 3000$~GeV may have brighter prospects: important
guidance will be provided by Run~2 of the LHC.

\begin{figure*}[htb!]
\begin{center}
\resizebox{9cm}{!}{\includegraphics{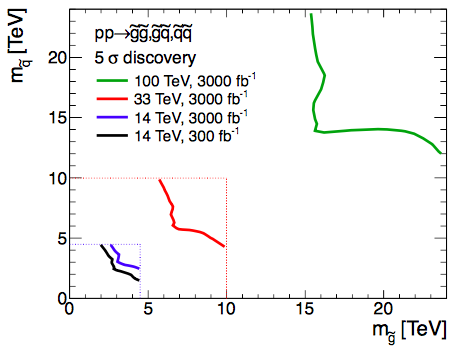}}
\end{center}
\vspace{-0.5cm}
\caption{\it Reaches for discovery of the gluino, ${\tilde g}$, and a generic squark, ${\tilde q}$, at the LHC operating at 14~TeV with
300 or 3000/fb of integrated luminosity, at a 33-TeV HE-LHC with 3000/fb, and at a 100-TeV collider with 3000/fb~\protect\cite{Snowmass}. 
}
\label{fig:100TeV}
\end{figure*}

Proton-proton colliders with centre-of-mass energies higher than the LHC are attracting
increased attention, notably the SppC project in China~\cite{SppC} and the FCC-hh project at CERN~\cite{FCC}.
The latter would be located in a circular tunnel $\sim 80$ to 100~km in circumference, and
aims at a centre-of-mass energy of 100~TeV.
Fig.~\ref{fig:100TeV} displays the reaches of various $pp$ colliders for searches for squarks
and gluinos~\cite{Snowmass}. With 3000/fb of integrated luminosity, a 100~TeV $pp$ collider would have a discovery reach $\sim 15$~TeV
for the squark or gluino mass, within simplified models. Increasing the target luminosity to 20,000/fb is currently under
discussion~\cite{Mangano}, which would increase the discovery reach to $\sim 20$~TeV.

How large might the squark and masses be? Of course, fine-tuning arguments would have favoured masses
$\lesssim 1$~TeV, but these arguments are notoriously imprecise and subjective. The DM
density provides more secure upper limits on sparticle masses in supersymmetric models, at the price of assuming
$R$ parity conservation and conventional adiabatic expansion of the Universe. As discussed previously,
LSP masses much above a TeV are possible only under exceptional conditions on the sparticle spectrum,
such as near-degeneracy between the LSP and the NLSP so that coannihilation is possible, or when the
LSP mass is close to half that of a direct-channel heavy Higgs resonance, so that LSP-LSP annihilation
is enhanced.

\begin{figure*}[htb!]
\begin{center}
\resizebox{8cm}{!}{\includegraphics{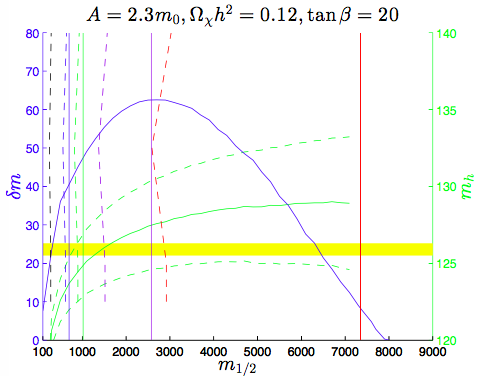}}
\resizebox{6.5cm}{!}{\includegraphics{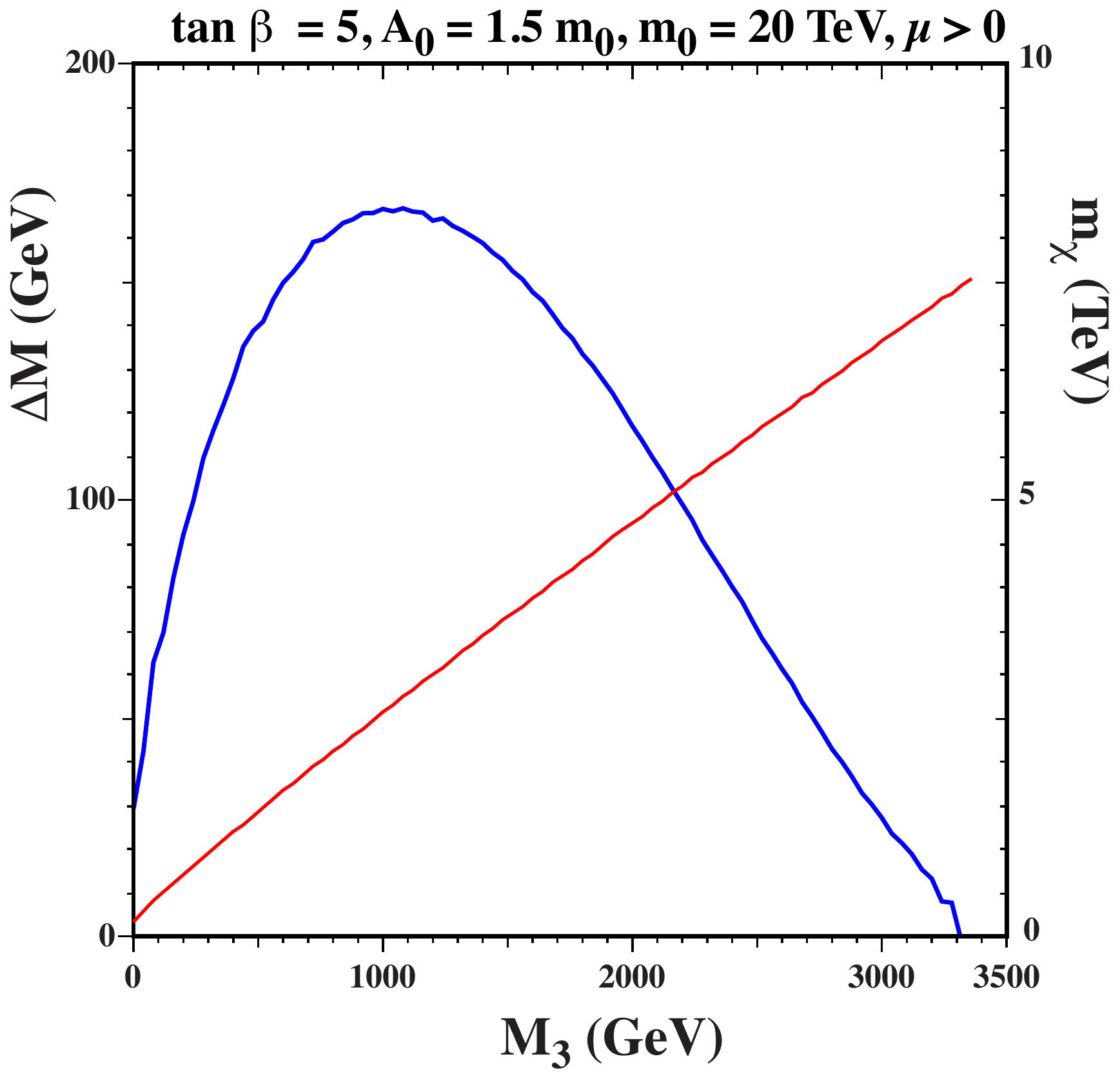}}
\end{center}
\vspace{-0.5cm}
\caption{\it Profiles of coannihilation strips corresponding to $\Omega_\chi h^2 = 0.120$ for (left panel) the 
lighter stop, ${\tilde t_1}$, in the CMSSM
for $A_0 = 2.3 m_0$ and $\tan \beta = 20$~\protect\cite{EOZ}, and (right panel) the gluino, ${\tilde g}$, in a variant of the CMSSM with non-universal gaugino masses
$M_! = M_2 \ne M_3$, $m_0 = 20$~TeV, $A_0 = 1.5 m_0$ and $\tan \beta = 20$~\protect\cite{EELO}, both with $\mu > 0$.
}
\label{fig:strips}
\end{figure*}

If the LSP is almost degenerate with the lighter stop, which is possible even in the CMSSM,
the LSP mass may be as large as $\sim 6.5$~TeV~\cite{EOZ}. This would seem to be within reach of a 100-TeV
collider. However, close to the end-point of the stop coannihilation strip, the mass difference
$m_{\tilde t_1} - m_\chi \to 0$, as seen in the left panel of Fig.~\ref{fig:strips}.
The black, blue, green, purple and red lines are particle exclusion reaches for particle searches with LHC at 8 TeV, 
300 and 3000/fb with LHC at 14 TeV, 3000/fb with HE-LHC at 33 TeV and 3000/fb with FCC-hh at 100 TeV, respectively. 
The solid lines are for generic MET searches, and the dashed lines are for dedicated stop searches.
Towards the end of the stop coannihilation strip, stop decays would not produce energetic jets and might be
difficult to detect directly. Nevertheless, in the CMSSM there would be other supersymmetric signatures
due, e.g., to the production and decays of gluinos and other squark flavours. Thus the CMSSM stop
coannihilation scenario should be accessible at a 100-TeV $pp$ collider, as shown by the solid red line in the left panel of Fig.~\ref{fig:strips}.

The LSP could be even heavier, $m_\chi \lesssim 8$~TeV, if it coannhilates with the gluino~\cite{ELO}. This is not
possible in the CMSSM, but is possible in models with non-universal gaugino masses: $M_1 \ne M_3$
at the input GUT scale, or in pure gravity mediation models with extra vector-like supermultiplets~\cite{EELO}. An
example of the former is shown in the right panel of Fig.~\ref{fig:strips}. In this case the direct production
of gluino pairs would produce energetic jets towards the tip of the gluino coannihilation strip, and the input scalar
mass $m_0 = 20$~TeV, rendering squarks difficult to detect. Moreover, larger values of $m_0$ are also
possible. Therefore the tip of the coannihilation strip where $m_\chi \sim m_{\tilde g} \sim 8$~TeV
could be challenging even for a 100-TeV $pp$ collider.

\section{Summary}

Rumours of the death of supersymmetry are greatly exaggerated. I consider that it is
still the best-motivated framework for TeV-scale physics, in view of its help in making the
fine-tuning of the mass hierarchy more natural, its help with grand unification and its r\^ole
in string theory. In addition, the LSP is still an excellent candidate for cold dark matter.
All that said, simple supersymmetric models with universal soft supersymmetry-breaking terms,
such as the CMSSM, etc., are under pressure. However, it is worth noting that much of the pressure
comes from $g_\mu - 2$, which is even more of an issue for the SM. Moreover, more general models 
within the pMSSM framework are quite healthy, since they can reconcile $g_\mu - 2$ with the
other constraints. Within the CMSSM, the NUHM1, the NUHM2 and the pMSSM10 we find
good prospects for detecting supersymmetry and possibly making some interesting measurements,
either at LHC Run 2 or via direct dark matter detection,
though there are no guarantees. We may well need a higher-energy $pp$ collider, either to make follow-up
studies of supersymmetry, or to discover it!

\end{document}